\documentclass[aps,prb,reprint,amsmath,amssymb,superscriptaddress,floatfix]{revtex4-2}

\usepackage[colorlinks,citecolor=blue,linkcolor=blue]{hyperref}
\usepackage{graphicx}
\usepackage{dcolumn}
\usepackage{bm}
\usepackage{upgreek}
\usepackage{txfonts}

\newcommand{\CNN}{Centre de Nanosciences et de Nanotechnologies, CNRS, Universit\'e Paris-Saclay, 91120 Palaiseau, France}
\newcommand{\UMPhy}{Unit\'e Mixte de Physique, CNRS, Thales, Universit\'e Paris-Saclay, 91767 Palaiseau, France} 
\newcommand{\IJL}{Institut Jean Lamour, CNRS, Universit\'e de Lorraine, 54011 Nancy, France}
\newcommand{\LMOPS}{Chair in Photonics, LMOPS EA 4423 Laboratory, CentraleSup{\'e}lec, Universit\'e de Lorraine, 57070 Metz, France}

\newcommand{\idc}{\ensuremath{I_\mathrm{DC}} }

\newcommand{\hx}{\ensuremath{H_\mathrm{x}} }

\begin{document}

\title{Nanocontact vortex oscillators based on Co$_2$MnGe pseudo spin valves}

\author{J{\'e}r{\'e}my L{\'e}tang}
\email{jeremy.letang@protonmail.com}
\affiliation{\CNN}
\author{Claudia de Melo}
\affiliation{\IJL}
\affiliation{\LMOPS}
\author{Charles Guillemard}
\affiliation{\IJL}
\author{Aymeric Vecchiola}
\affiliation{\UMPhy}
\author{Damien Rontani}
\affiliation{\LMOPS}
\author{S{\'e}bastien Petit-Watelot}
\affiliation{\IJL}
\author{Myoung-Woo Yoo}
\affiliation{\CNN}
\affiliation{\LMOPS}
\author{Thibaut Devolder}
\affiliation{\CNN}
\author{Karim Bouzehouane}
\affiliation{\UMPhy}
\author{Vincent Cros}
\affiliation{\UMPhy}
\author{St{\'e}phane Andrieu}
\affiliation{\IJL}
\author{Joo-Von Kim}
\email{joo-von.kim@c2n.upsaclay.fr}
\affiliation{\CNN}

\date{21 June 2021}

\begin{abstract}
We present an experimental study of vortex dynamics in magnetic nanocontacts based on pseudo spin valves comprising the Co$_2$MnGe Heusler compound. The films were grown by molecular beam epitaxy, where precise stoichiometry control and tailored stacking order allowed us to define the bottom ferromagnetic layer as the reference layer, with minimal coupling between the free and reference layers. 20-nm diameter nanocontacts were fabricated using a nano-indentation technique, leading to self-sustained gyration of the vortex generated by spin-transfer torques above a certain current threshold. By combining frequency- and time-domain measurements, we show that different types of spin-transfer induced dynamics related to different modes associated to the magnetic vortex configuration can be observed, such as mode hopping, mode coexistence and mode extinction appearing in addition to the usual gyration mode.
\end{abstract}

\maketitle

\section{Introduction} 
Spin-torque nano-oscillators (STNOs) are spintronic devices in which applied currents drive self-sustained magnetization oscillations, which provide a basis for electrical microwave oscillators on the nanoscale. They are promising for a number of different applications, such as rf communications~\cite{choi_spin_2014, ruiz-calaforra_frequency_2017}, field sensing~\cite{tulapurkar_spin-torque_2005, jenkins_spin-torque_2016} and neuro-inspired computing~\cite{torrejon_neuromorphic_2017, romera_vowel_2018, tsunegi_physical_2019, williame_chaotic_2019}. An important challenge is to reduce the operating currents and improve signal-to-noise ratios. In this light, Heusler compounds~\cite{heusler_friedrich_uber_1903, graf_simple_2011} are promising materials to meet these challenges. The near half-metallicity of these materials result in ultralow damping  compared to standard ferromagnets~\cite{andrieu_direct_2016} and a large spin polarization~\cite{kubler_understanding_2007}, which reduces the threshold of spin-transfer torques and results in a larger magnetoresistive ratio in spin valve structures. This choice has been explored in vortex-based systems involving Co$_2$Fe$_x$Mn$_{1-x}$Si~\cite{yamamoto_magnetic_2016, yamamoto_vortex_2016, seki_size_2018}, where a large output power of 10.2 nW~\cite{yamamoto_vortex_2016} and high quality factor of 5000~\cite{seki_size_2018} have been reported.

In this paper, we present an experimental study of nanocontact vortex oscillators based on pseudo spin valves using Co$_2$MnGe Heusler compounds. These are members of the broader class of materials with the composition Co$_2$Mn$Z$, where $Z$ are different III B, IV B or V B elements, which have been studied in detail recently~\cite{andrieu_direct_2016, guillemard_polycrystalline_2019, guillemard_ultralow_2019}. In particular, it has been demonstrated that Co$_2$MnSi and Co$_2$MnGe exhibit a combination of the lowest Gilbert damping, $\alpha$, and highest spin polarization, $P$, among this class~\cite{guillemard_ultralow_2019}. In particular, ferromagnetic resonance experiments have shown that $\alpha = 4.6 \times 10^{-4}$ for Co$_2$MnSi and $\alpha = 5.3 \times 10^{-4}$ for Co$_2$MnGe at 290 K~\cite{guillemard_ultralow_2019}, and both have close to the maximum polarization of $P=1$, which are close to theoretical predictions \cite{de_groot_new_1983, ishida_search_1995, picozzi_mathrmco_2mathrmmnx_2002, kandpal_calculated_2007, liu_origin_2009}. Lattice parameter mismatch is smaller between Co$_2$MnGe and Au than between Co$_2$MnSi and Au, which is used as spacer. We report here some results obtained on Co$_2$MnGe-based spin valves.

A particular magnetic structure that underpins a class of spin-torque nano-ocillators is the magnetic vortex. In the nanocontact geometry, for example, the magnetic vortex represents a metastable state that is nucleated when the Zeeman potential associated with applied currents flowing through the nanocontact becomes sufficiently strong. As described by the Thiele equation~\cite{thiele_steady-state_1973}, the vortex core gyrates around the nanocontact above a certain current threshold, whose motion is determined by the balance between the restoring force of the Zeeman confining potential, Gilbert damping, and spin torques. This steady-state gyration leads to a periodic variation of the magnetization in the nanocontact area, which can be detected electrically through resistance variations in a magnetoresistive stack such as spin valves. The sense of gyration is determined by the core polarity \cite{thiele_steady-state_1973}.

A unique feature of nanocontact vortex oscillators is the onset of periodic core reversal above a critical current~\cite{petit-watelot_commensurability_2012, devolder_chaos_2019, letang_modulation_2019, yoo_pattern_2020}. This core reversal occurs concurrently with the steady-state gyration around the nanocontact, giving rise to self-phase locked and chaotic states. Core reversal involves the creation and annihilation of vortex-antivortex pairs~\cite{van_waeyenberge_magnetic_2006}, where the latter results in the emission of spin wave bursts~\cite{hertel_exchange_2006}. This dynamics is driven primarily by currents flowing in the film plane (CIP), resulting in spin-transfer torques of the Zhang-Li form~\cite{zhang_roles_2004} rather than the Slonczewski torques~\cite{slonczewski_current-driven_1996} associated with currents perpendicular to the film plane in nanopillar geometries~\cite{pribiag_magnetic_2007, dussaux_large_2010, Locatelli:2011hw}. Therefore, it is interesting to enquire whether the low damping afforded by Heusler materials can affect the gyration process significantly as well as the synchronization process between gyrotropic motion and the core reversal since spin-wave relaxation times greatly exceed the typical timescale of the gyration and core-reversal periods. As we show here, the current-driven dynamics in Heusler-based pseudo spin valves can be quite rich and differ qualitatively from the behavior observed in permalloy-based systems. Some of the behaviors we observe might be interpretable as the result of coupled vortex dynamics in the free and reference layers.

\begin{figure*}
\centering\includegraphics[width=\textwidth]{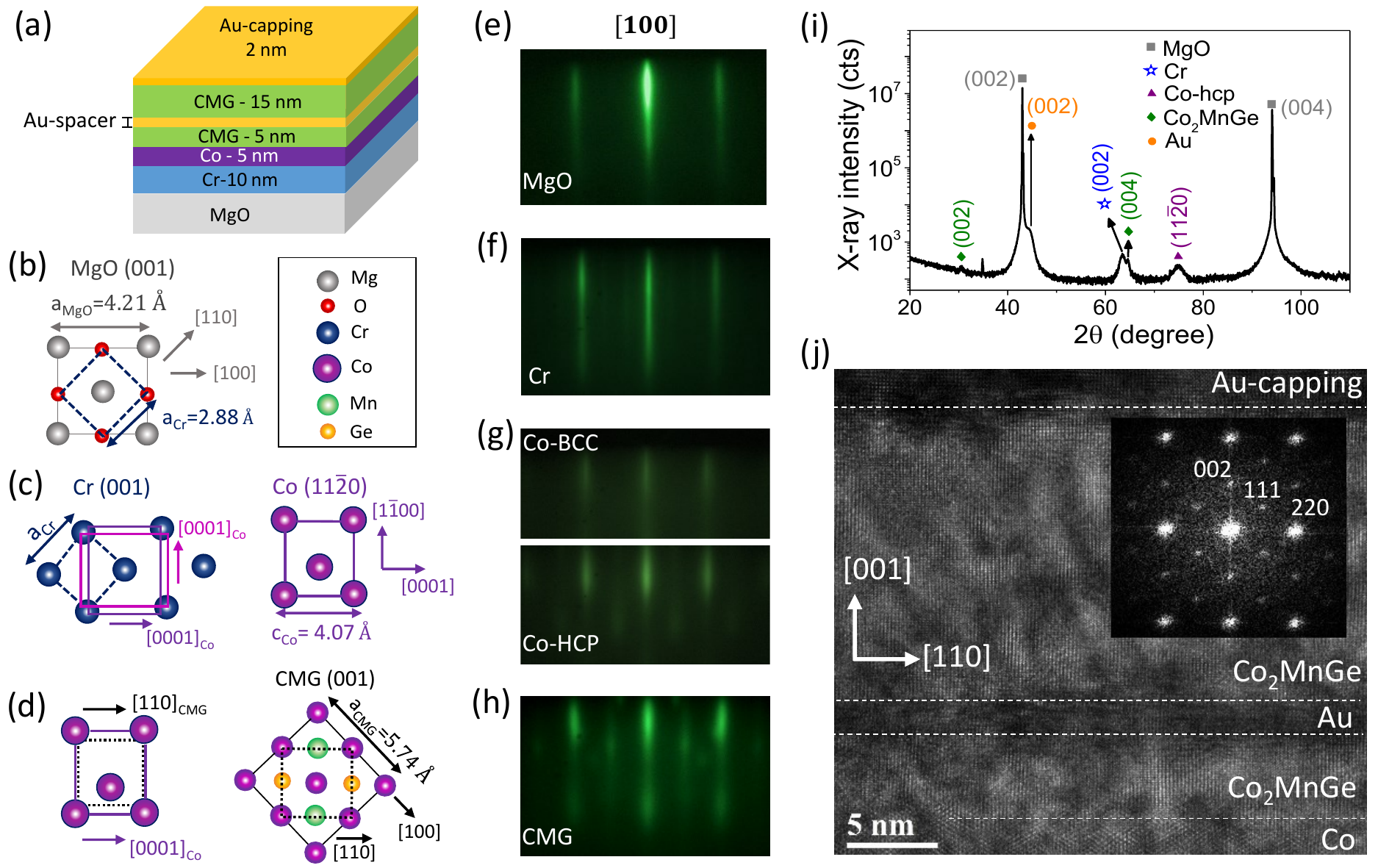}
\caption{(a) Schematic of the pseudo spin valve stack. (b) Epitaxial relationships between the Cr layer and the MgO substrate. (c) Epitaxial relationship between the Co and Cr layers (left), and schematic of the HCP Co $(11\overline{2}0)$ planes (right). (d) Epitaxial relationship between the Heusler Co$_2$MnGe film (noted CMG) and HCP Co (left), and schematic of CMG (001) planes (right). (e)$-$(h) RHEED patterns obtained for each layer of the stack along the [100] azimuth of the MgO substrate. (i) X-ray diffractogram of the pseudo spin valve. (j) High-resolution TEM micrograph of the pseudo spin valve stack. Fourier transform (FT) pattern performed to the Co$_2$MnGe free layer along the [110] axis is displayed at the inset. }
\label{Figure1structure}
\end{figure*}

This paper is organized as follows. In Sec. II, we describe the epitaxial film growth, along with the structural and magnetic characterization of the Co$_2$MnGe films using x-ray diffractometry, \emph{in situ} reflection high-energy electron diffraction (RHEED), vibrating sample magnetometry (VSM), and broadband ferromagnetic resonance (FMR). In Sec. III, we describe the frequency-domain analyses of the electrical power spectra. This is followed by a description of the time-domain analyses in Sec. IV, where mode stability is discussed. A discussion and concluding remarks are given in Sec. V.

\section{Film growth and sample fabrication}

The pseudo spin valves based on Heusler compounds were grown by molecular beam epitaxy (MBE) on (001)-oriented MgO substrates. The multilayer stack studied is composed of Cr (10 nm) / Co (5 nm) / Co$_2$MnGe (5 nm) / Au (1$-$10 nm) / Co$_2$MnGe (15 nm) / Au-capping (2 nm), as schematically depicted in Fig.~\ref{Figure1structure}(a). The stoichiometry of the Co$_2$MnGe films was accurately controlled by calibrating the atomic fluxes of each element through a quartz microbalance located at the sample position ~\cite{guillemard_issues_2020}. The 15-nm-thick Co$_2$MnGe corresponds to the free layer, while the Co(5 nm)/Co$_2$MnGe (5 nm) composite bilayer plays the role of the reference layer in the pseudo spin valve as it possesses a larger coercive field compared with the Co$_2$MnGe layer. In MBE-grown structures, it is often difficult to achieve a reference magnetic layer as a bottom electrode due to epitaxy constraints. Here, we succeeded in this respect by growing the 5-nm Co$_2$MnGe film biased through exchange coupling with an epitaxial Co hard layer, as explained in more detail in the following. 

First, a 10-nm-thick Cr buffer layer was grown on the MgO substrate to promote the epitaxial growth of hexagonal closest packed (hcp) Co which has a high magnetic anisotropy compared with Co$_2$MnGe. X-ray diffractograms and RHEED patterns [Figs.~\ref{Figure1structure}(e) and \ref{Figure1structure}(i)] confirm the epitaxial relationship: $[110]\;(001)$ Cr $\|$ $[100]\;(001)$ MgO, where Cr $[100]$ axis is 45$^{\circ}$ rotated from MgO $[100]$ axis [see schematic in Fig. \ref{Figure1structure}(b)]. The Co layer starts to grow on Cr with the body-centred cubic (bcc) crystal structure and relaxes to its hcp structure after a critical thickness around four atomic planes~\cite{andrieu_spectroscopic_2014}, as indicated by the appearance of new streaks in the RHEED pattern of Fig. \ref{Figure1structure}(g). RHEED patterns [Fig. \ref{Figure1structure}(g)] and x-ray diffractograms [Fig. \ref{Figure1structure}(i)] are consistent with the epitaxial relationship of $[0001]\;(11\overline{2}0)$ Co $||$ $[110]\;(001)$ bcc Cr. The same RHEED pattern was obtained by turning the sample 90$^{\circ}$, which means that Co $[0001]$ direction is a fourfold rotational symmetry axis. This is consistent with considering two domains 90$^{\circ}$ rotated in-plane, that is, the Co $[0001]$ azimuth parallel to either the $[110]$ or $[1\overline{1}0]$ Cr azimuths, as sketched in Fig. \ref{Figure1structure}(c). A similar bicrystal epitaxial relation has been reported for several Co/Fe and Co/Cr layers deposited by MBE~\cite{nakamura_epitaxial_1993, wang_structural_2007,popova_epitaxial_2002}. Co$_2$MnGe was observed to epitaxially grow on the relaxed Co layer with the epitaxial relationship  $[110]\;(001)$ Co$_2$MnGe $\|$ $[0001]\; (11\overline{2}0)$ hcp Co [Fig. \ref{Figure1structure}(d), left], as confirmed by the RHEED pattern performed at the film surface [Fig.~\ref{Figure1structure}(h)]. Gold was chosen for the spacer layer first because epitaxial and continuous Au films were obtained on Co$_2$MnGe~\cite{guillemard_ultralow_2019} and, second, because Co$_2$MnGe was also observed to grow epitaxially on Au. Both Co$_2$MnGe films are consistent with a chemically ordered L2$_1$ structure, as indicated by x-ray diffraction measurements, by the presence of half streaks in the RHEED patterns taken along the Co$_2$MnGe $[110]$ azimuth [see, for example, Fig. \ref{Figure1structure}(h)] and confirmed by high-resolution transmission electron microscopy [Fig. \ref{Figure1structure}(j)].

\begin{figure}
\centering\includegraphics[width=8.5cm]{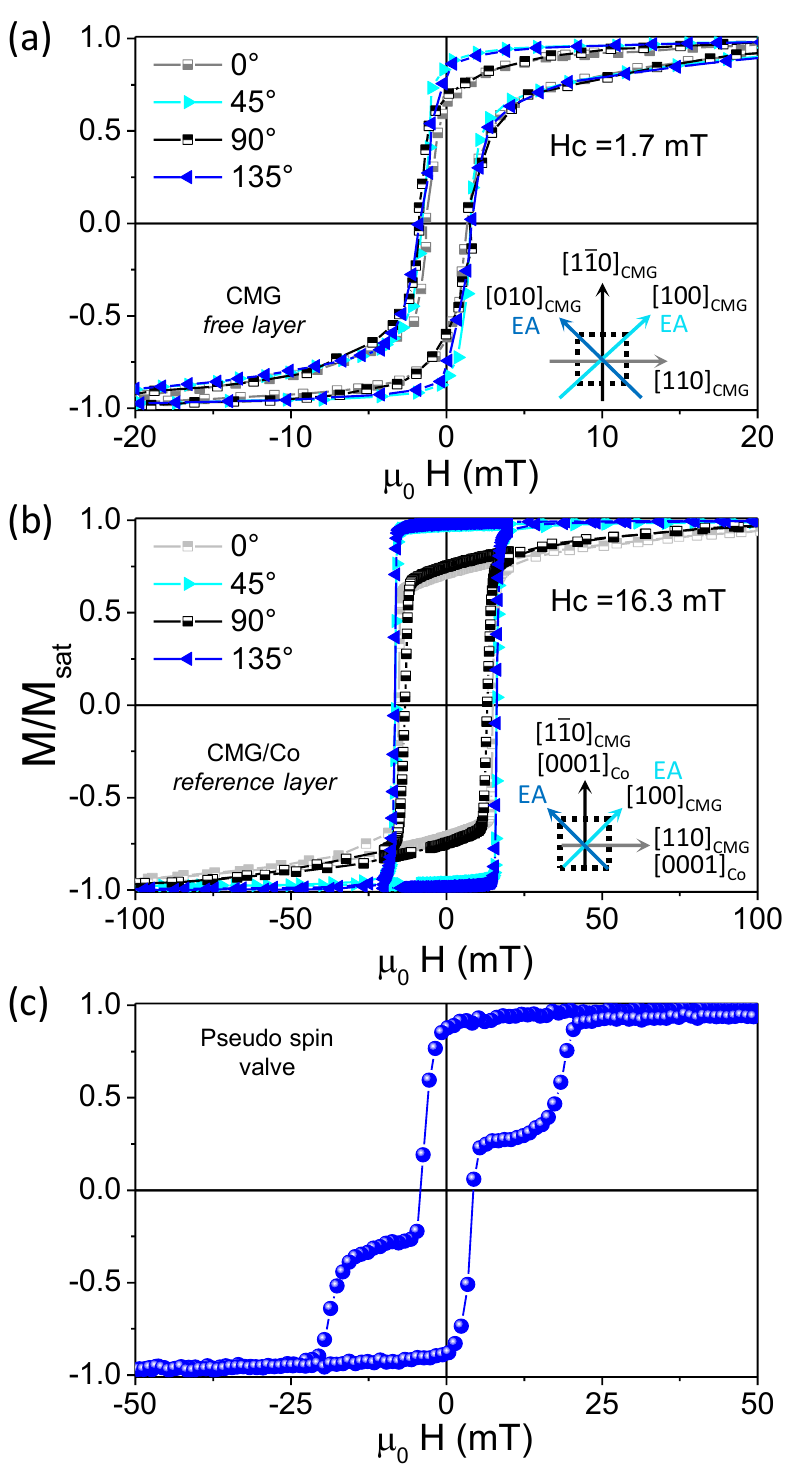}
\caption{In-plane hysteresis loop at different angles for (a) the free layer: CMG / MgO,  (b) the reference layer: CMG / Co / Cr / MgO, and (c) the pseudo spin valve with Au-spacer layer of 2 nm (EA meaning easy axis). Horizontal scales are different between figures. }
\label{Figure2vsm}
\end{figure}

The magnetic coupling between the Co and bottom Co$_2$MnGe layers is confirmed by hysteresis loop measurements. In Fig.~\ref{Figure2vsm}, the in-plane M-H hysteresis loops of the reference layer Co$_2$MnGe (5 nm) / Co (5 nm) / Cr (10 nm) / MgO [Fig.~\ref{Figure2vsm}(b)] is compared with the hysteresis loop of a single Co$_2$MnGe soft magnetic layer (15 nm) / MgO [Fig.~\ref{Figure2vsm}(a)]. The magnetic field is applied in plane along the main crystallographic axes of MgO (MgO $[001]$ axis $\Rightarrow$ 0$^{\circ}$), as sketched in the inset of Fig.~\ref{Figure2vsm}. The soft Co$_2$MnGe magnetic layer [Fig.~\ref{Figure2vsm}(a), coercivity of 1.6 mT] shows an in-plane four-fold magnetic anisotropy with easy axes at $[100]$ and $[010]$, which corresponds to the curves at 45$^{\circ}$ and 135$^{\circ}$ since the $[100]$ axis of Co$_2$MnGe is rotated by 45$^{\circ}$ about the $[100]$ axis of MgO [see inset in Fig.~\ref{Figure2vsm}(a)]. The coercivity enhancement  in the reference layer (16.3 mT) is evidenced in Fig.~\ref{Figure2vsm}(b), which confirms the strong exchange coupling between Co$_2$MnGe and Co layers. It is well known that bulk hcp Co has a uniaxial magnetocrystalline anisotropy along its $[0001]$ axis, denoted here as the $c$ axis. According to the epitaxial relationship of the two Co domains here, the $c$ axes are in plane and 90$^{\circ}$ rotated from one domain to the other, leading to a four-fold magnetic anisotropy, with easy axes rotated 45$^{\circ}$ from the Co $[0001]$ direction~\cite{gu_fourfold_1995}. This finally leads to easy axes in the Co/Co$_2$MnGe stack similar to the ones in Co$_2$MnGe [see schematic at the inset of Fig.~\ref{Figure2vsm}(b)]. This explains the four-fold magnetic anisotropy of the reference layer, with easy axes along $[100]$ and $[010]$ Co$_2$MnGe directions, and the improved squareness of the hysteresis loop along these axes, compared with the single Co$_2$MnGe layer.

The choice of the Au-spacer thickness represents a compromise between conduction electron scattering and interlayer exchange coupling (IEC) between the free and reference layers through Au. In this paper, Au-spacer layers of different thicknesses ranging from 1 to 10 nm were explored. A well-defined double-stepped hysteresis loop was obtained from 2 nm, which means that this thickness is enough to keep the free and reference layers magnetically decoupled [see Fig.~\ref{Figure2vsm}(c)].

The stacks with 2-nm Au spacers were used to fabricate nanocontact oscillators. We defined $70 \times 30$ $\mu$m$^2$ mesas and an indentation back electrode wire in the stack down to substrate with argon ion beam etching. We first define a thin insulating resist (SU8) 40 nm thick, with a surface area of 40$\times$40 $\mu$m$^2$ over the mesa. This layer represents the ``indentation mask'' through which the nano-indentation is performed. A thick resist (1-$\mu$m-thick SU8) was deposited to provide a good insulation between top and base electrodes over large areas. An aperture of 20$\times$20 $\mu$m$^2$ is opened in the thick resist above the thin resist, defining the ``indentation zone''. A third resist is defined by optical lithography for top electrode lift-off. A nano-hole is thus created in the thin resist by using a conducting atomic force microscope tip as described in Ref.~\cite{bouzehouane_nanolithography_2003}. Samples were placed in a gentle reactive plasma (15 W) in 75\% Ar / 25\% O$_2$ gas mixture to clean up any remaining resist at the bottom of the nano-indented hole, i.e., at the surface of the stack, and to finely tune the nanocontact diameter. Following this,  tantalum and gold were deposited to form the nanocontact and top electrodes. This process enabled us to realize nanocontacts with radii of 20 nm. The back electrode wires used for electrical control of the AFM tip is finally cut with a diamond spear.

We assume that these properties vary little between layers on the substrate and in the stack. Under a magnetic field loop, we measure at best for the full stack with nanocontact a magnetoresistance (MR) ratio of $0.12\pm0.03 \%$. A single 20-nm-layer of Co$_2$MnGe has an anisotropic magnetoresistance (AMR) ratio of 0.19\%, which is comparable to the MR ratio of the full stack. However, the indentation process quality impacts the nanocontact quality, which can change the MR ratio.

\section{Frequency domain analysis}
\begin{figure}
\centering\includegraphics[width=8cm]{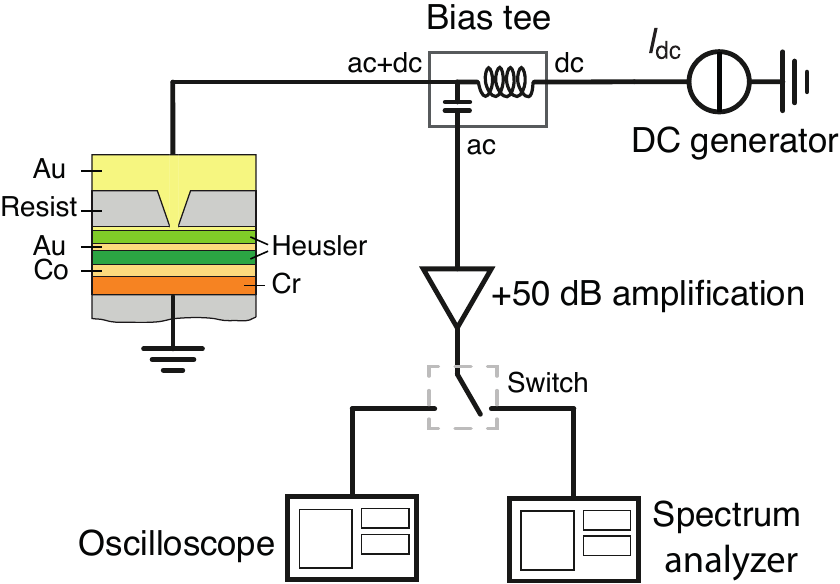}
\caption{Electrical circuit for frequency- and time-domain measurements.}
\label{circuit}
\end{figure}

The microwave frequency measurement setup is illustrated in Fig.~\ref{circuit}.
A bias tee separates the DC current that is applied to the nanocontact and the high-frequency components that are generated from spin-torque driven dynamics of the magnetic vortex. The high-frequency branch is amplified by +50 dB before reaching a switch that directs the signal either to the spectrum analyzer or the oscilloscope. An in-plane magnetic field was applied on the sample.

As in previous studies of nanocontact vortex oscillators~\cite{petit-watelot_commensurability_2012, devolder_chaos_2019, letang_modulation_2019, yoo_pattern_2020}, it is necessary to nucleate a vortex (nominally in the free magnetic layer) as this state is not necessarily the ground state of the magnetization. This is achieved by applying a large DC current, here 9 or 10 mA which is toward the upper limit before sample degradation, while sweeping an in-plane magnetic field. Simulations have shown previously in samples~\cite{petit-watelot_commensurability_2012, devolder_chaos_2019, letang_modulation_2019, yoo_pattern_2020} with a different material that as the magnetization reverses, a domain wall sweeps through the free layer and as it crosses the nanocontact region, a vortex is nucleated through the competition between the current-induced Oersted field and spin-transfer torques. This nucleation process can result experimentally in slightly different micromagnetic configurations from one nucleation event to the next, which results in power spectra that can vary qualitatively.

\begin{figure}
\centering\includegraphics[width=8cm]{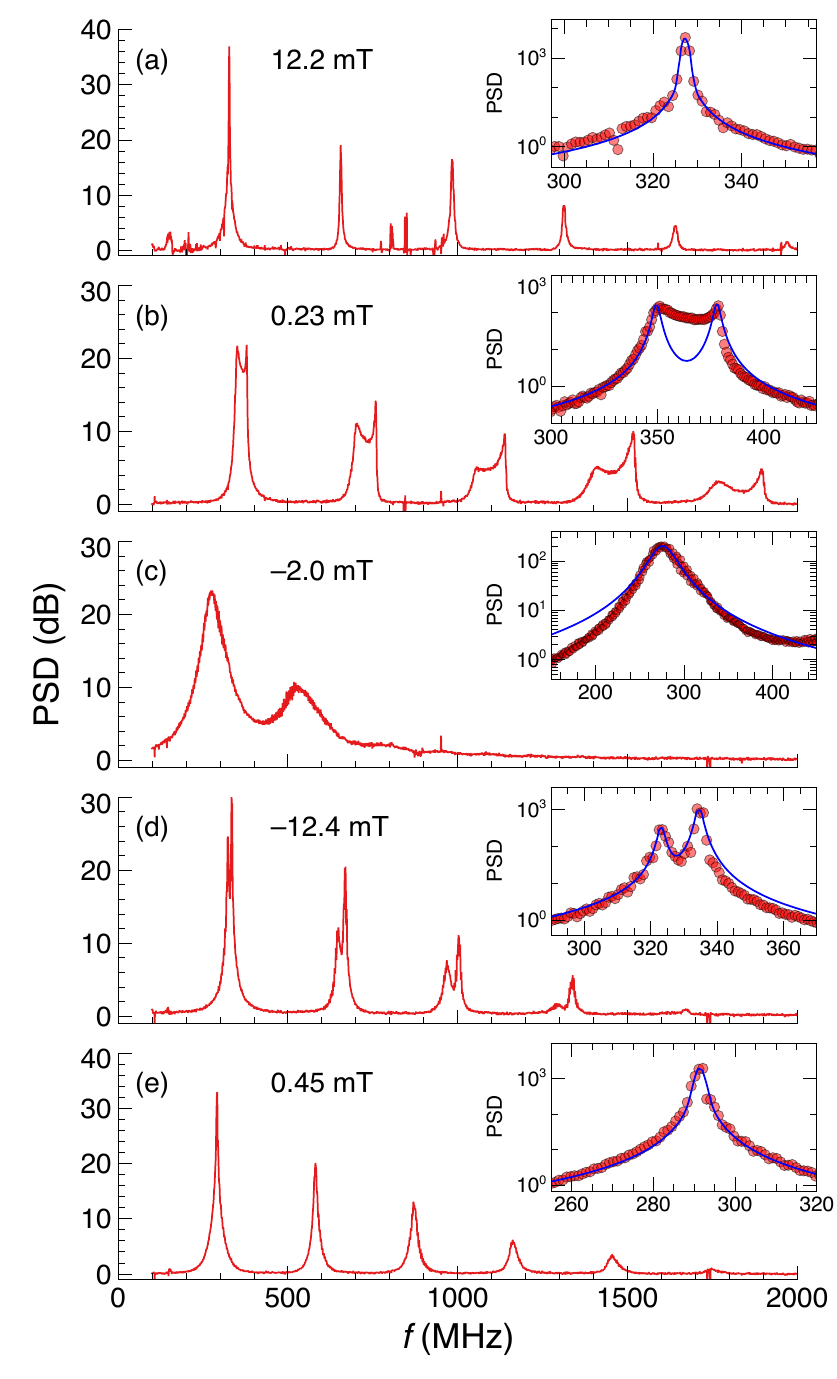}
\caption{Power spectral density (PSD) of magnetoresistance oscillations at $I_\mathrm{DC} = 8.5$ mA for nucleation events under different applied in-plane fields. (a) Single mode with narrow spectral peaks and higher order harmonics at $\mu_0 \hx = 12.2$ mT, (b) double frequency mode at $\mu_0 \hx = 0.23$ mT, (c) broad asymmetric peaks at $\mu_0 \hx = -2.0$ mT, (d) double frequency mode at $\mu_0 \hx = -12.4$ mT, and (e) single mode at $\mu_0 \hx = 0.45$ mT. The PSD is given in relative power with respect to a reference signal that has been subtracted from the curves displayed. The insets show fits to the fundamental frequency peak using a pseudo-Voigt function for (a) and (e), a double Lorentzian for (b) and (d), and single Lorentzian for (c).}
\label{heuslercurves}
\end{figure}

Examples of the different power spectra observed are presented in Fig.~\ref{heuslercurves} for a nominal applied current of 8.5 mA, which establishes a common Zeeman energy potential associated with the current-induced Oersted-Amp{\`e}re fields for the vortex gyration~\cite{mistral_current-driven_2008}, but under different nucleation and applied field conditions.
For these spectrum analyzer measurements, we used a resolution bandwidth of 500 kHz, a video bandwidth of 50 kHz, a sweep time of 10 s per spectrum, and 2000 points were acquired per spectrum. The power spectral density (PSD) of the measured MR oscillations shown is obtained after subtracting off a reference signal without spin-torque driven oscillations; we give it in dB as here it represents a relative power above this reference signal. We can observe well-defined peaks in the sub-GHz regime, which is consistent with vortex gyration around the nanocontact with radii in the range of 100$-$200 nm.

Figure~\ref{heuslercurves}(a) represents a case where the measurement was made under an in-plane applied field of 12.2 mT, with the frequency of the main peak at $f_0 = 327.3$ MHz and higher order harmonics up to $5f_0$ are visible in the frequency range displayed. Some parasitic noise is also observed, in particular, between the second and third harmonics. However, these peaks do not vary with applied current and can be attributed to artifacts related to external parasitic sources such as telecommunication networks. The spectral line of the fundamental frequency is in the range of 1 MHz, which is comparable to the resolution bandwidth used. As such, we find that the spectral line shape is best described by the pseudo-Voigt function
\begin{multline}
S({\nu}) = A \, \Biggl[ \eta \, \frac{1}{\sigma \sqrt{2\pi} } \exp\left(-\frac{(\nu-\nu_0)^2}{2\sigma^2} \right) \\
+ (1-\eta) \, \frac{1}{\pi}\frac{\Gamma/2}{(\nu-\nu_0)^2+(\Gamma/2)^2}  \Biggr],
\label{eq:pseudovoigt}
\end{multline}
with
\begin{equation}
\sigma = \frac{\Gamma}{2 \sqrt{2 \ln 2}},
\end{equation}
which represents a linear combination of a Gaussian line shape and a Lorentzian line shape, and approximates the Voigt function that describes the convolution between the Gaussian and Lorentzian functions. Here, $A$ represents the amplitude, $\nu_0$ is the central frequency, $\Gamma$ is the full width at half maximum (which we refer to simply as the linewidth), and $0 \leq \eta \leq 1$ describes the relative contribution of the (normalized) Gaussian line shape with respect to the (normalized) Lorentzian profile. The inset of Fig.~\ref{heuslercurves}(a) shows that the pseudo-Voigt function [Eq.~(\ref{eq:pseudovoigt})] accounts very well for the spectral line shape of the fundamental frequency peak, where we find a linewidth of $\Gamma = 1.5$ MHz, which is similar to characteristics seen in permalloy-based nanocontact vortex oscillators~\cite{devolder_chaos_2019}, and $\eta = 0.77$, which is consistent with the fact that the linewidth is close to the resolution bandwidth.

Indeed, qualitatively different behavior can be seen under other nucleation conditions, such those shown in Figs.~\ref{heuslercurves}(b)$-$\ref{heuslercurves}(d). In Fig.~\ref{heuslercurves}(b), we observe a clear double-peak structure with high spectral power between the two maxima and a strong asymmetry across all the visible harmonics. Here, the power spectrum cannot be accounted for by a double Lorentzian fit as shown in the inset (we can neglect the Gaussian contribution because the features are much larger than the resolution bandwidth), where we can see the strongest discrepancy occurs in the region between the two maxima. In Fig.~\ref{heuslercurves}(c), we can observe a broad asymmetric triangular-shaped spectral line with a comparatively large linewidth of 32 MHz, which is suggestive of a chaotic process~\cite{devolder_chaos_2019}. An asymmetric double-peak structure can be seen Fig.~\ref{heuslercurves}(d), which, in contrast to the case shown in Fig.~\ref{heuslercurves}(b), can in fact be described by a double Lorentzian profile where the fitted central frequencies are 323.2 and 334.5 MHz, with linewidths of 2.5 and 2.4 MHz, respectively.

Finally in Fig.~\ref{heuslercurves}(e), we observe a similar case to Fig.~\ref{heuslercurves}(a) but with a lower in-plane applied field of 0.45 mT, where the fundamental peak appears at a lower frequency of 291.4 MHz with a broader spectral line. Again, this case can be well described by the pseudo-Voigt fit from which we obtain $\Gamma = 2.5$ MHz and $\eta = 0.4$. We note that the larger contribution from the Lorentzian profile is consistent with the larger linewidth, where convolution effects are less important.

\begin{figure}
\centering\includegraphics[width=8cm]{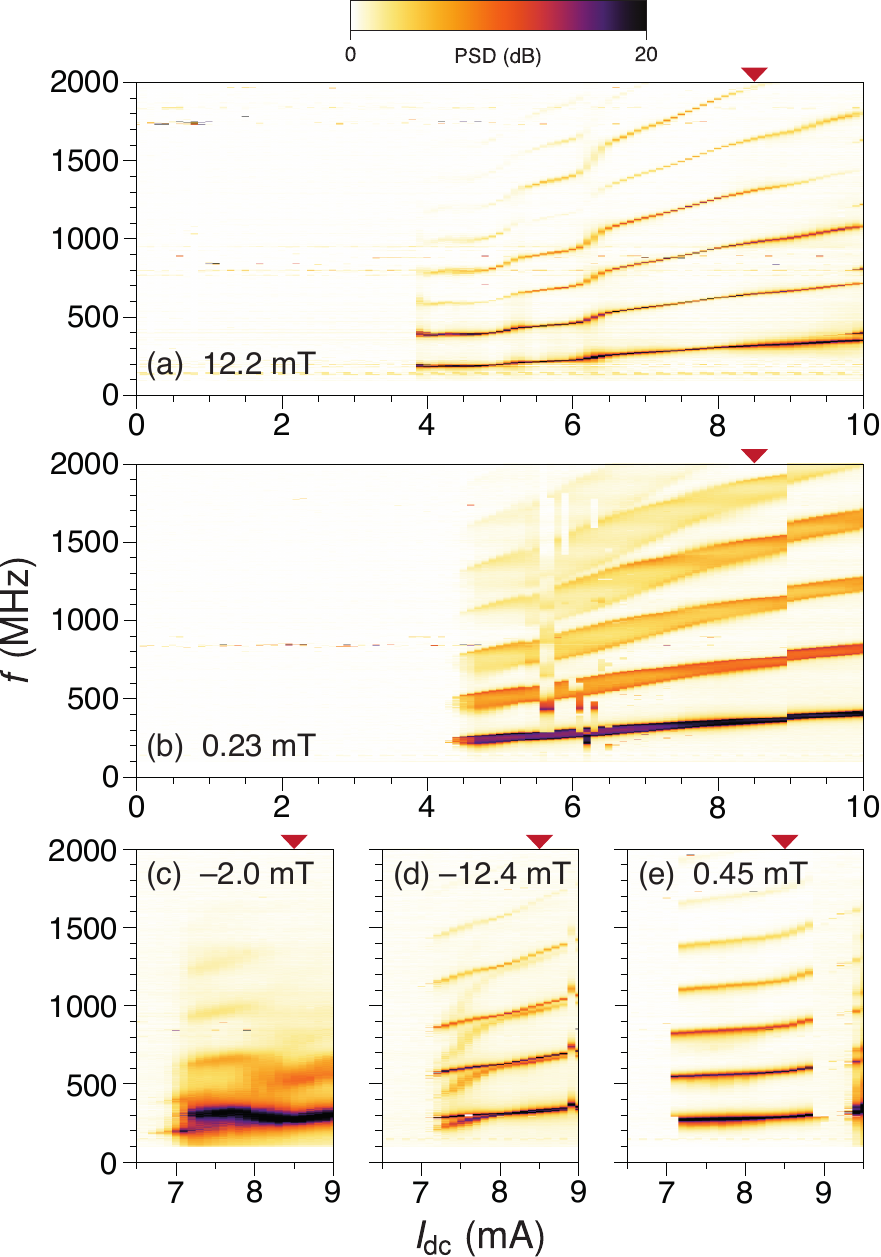}
\caption{Power spectral density (PSD) maps as a function of DC current for given magnetic field and different nucleation events. Spectra are measured every 0.1 mA. (a) Narrow signal at $\mu_0 \hx = 12.2$ mT, (b) two-frequency mode at $\mu_0 \hx = 230$ $\mu$T, (c) chaotic signal at $\mu_0 \hx = -2.0$ mT, (d) two-frequency mode at $\mu_0 \hx = -12.4$ mT, (e) narrow signal at $\mu_0 \hx = 450$ $\mu$T. The red arrows indicate the corresponding spectra shown in Fig. \ref{heuslercurves}.}
\label{heuslerpsd}
\end{figure}

In Fig.~\ref{heuslerpsd}, we present color maps of the current dependence of these power spectra, which exhibit a richer behavior than permalloy standard nanocontact oscillators~\cite{letang_modulation_2019}.
The color map represents a density plot in which the power spectra are shown as a function of applied current by assigning a color value to the amplitude of the peaks shown in in Fig.~\ref{heuslercurves}. Each of the five maps in Fig.~\ref{heuslerpsd} were acquired after a single nucleation event, where an example from each map is given in Fig.~\ref{heuslercurves}. Overall, we can observe an increase in the frequency of the main spectral line (along with its harmonics) as a function of applied current, which is consistent with behavior reported previously for nanocontact vortex oscillators. The spectra are acquired for descending current after successful nucleation. While gyration can be observed for currents down to around 4 mA in certain cases [Figs.~\ref{heuslerpsd}(a) and \ref{heuslerpsd}(b)], the signal can disappear at a higher threshold current of 7 mA, as shown in Figs.~\ref{heuslerpsd}(c)$-$\ref{heuslerpsd}(e). Thin horizontal lines in the color maps represent artifacts arising from parasitic electromagnetic signals, such as cellular telephone networks or WiFi signals, which are independent of the current and persist in the absence of the oscillator signal.

These color maps highlight some of the different qualitative behaviors observed. Figure~\ref{heuslerpsd}(a) represents the canonical case of nanocontact vortex oscillations with pure gyration without core reversal~\cite{pufall_low-field_2007, mistral_current-driven_2008, keatley_direct_2016}, where the frequency of the spectral peaks vary monotonically as a function of applied current. Some steplike changes are seen in this variation, a feature that has been reported in other STNOs and attributed to the transition between different oscillatory mode profiles~\cite{krivorotov_large-amplitude_2007}. The disappearance of the signal below 4 mA can either result from the annihilation of the vortex or pinning, whereby the spin-transfer torques do not allow the core to escape some local potential. The fact that this lower threshold varies between nucleation events suggests that this pinning is, in part, determined by the particular micromagnetic state present.

The current dependence of the double-peaked structure in Fig.~\ref{heuslercurves}(b) is given in Fig.~\ref{heuslerpsd}(b), where a distinctive bandlike structure can be observed over the entire current interval over which the oscillations are present. As we indicate above, a double-Lorentzian fit does not adequately describe the observed spectra and provides a particularly poor description of the large levels of power observed in the frequency interval between the two peaks. This feature is particularly apparent in the color map in Fig.~\ref{heuslerpsd}(b). The spectral profile is reminiscent of the power spectrum of a chirp signal, where the frequency varies in time between $f_1$ and $f_2$, the positions of the two peaks. However, we have not identified a physical mechanism that would produce such a signal in our device.

Figure~\ref{heuslerpsd}(c) provides an example of broad spectral lines, such as the PSD presented in Fig.~\ref{heuslercurves}(c), which might be interpreted as a signature of chaotic behavior. Indeed, the vortex gyration does not result in a sharp spectral line as in Fig.~\ref{heuslerpsd}(a) that is typically seen for simple vortex gyration around the nanocontact. Chaos in nanocontact systems studied previously results from competing behaviors involving gyration and periodic reversals of the vortex core~\cite{petit-watelot_commensurability_2012, devolder_chaos_2019, letang_modulation_2019, yoo_pattern_2020}. We do not see any evidence here of core reversal, which is typically characterized by the appearance of additional modulation sidebands in the power spectra. This suggests that another mechanism involving transitions in the micromagnetic state might be present. We note that hints of a double-mode structure appear at lower currents (i.e., around 7 mA) before the onset of the broad excitation peak as the current is increased, which suggests that the dynamics here results from the interaction between two oscillatory modes.

In Fig.~\ref{heuslerpsd}(d), we can see two distinct frequencies that are not related harmonically which appear to merge as the current is increased. In contrast to the case shown in Fig.~\ref{heuslercurves}(b), the double-peak structure here can be well described by a double Lorentzian fit, as illustrated in Fig.~\ref{heuslercurves}(d). Such a double-frequency mode has been observed in other STNOs \cite{krivorotov_large-amplitude_2007, pribiag_long-timescale_2009, kuepferling_two_2010, eggeling_low_2011, wang_multiple-mode_2011, keatley_imaging_2017, zhang_tunable_2017} and can either result from the hopping between two frequencies or from the simultaneous emission of two frequencies. We cannot distinguish between these two scenarios from frequency domain measurements alone and further discussion on this point will be given in the following section.

Finally, in Fig~\ref{heuslerpsd}(e), we observe a behavior consistent with the pure gyration regime but with the appearance of a microwave-quiet region within a small current interval between 8.9 and 9.3 mA. Such an extinction has been observed in other studies~\cite{martin_tunability_2013, soucaille_nanocontact_2017} but its mechanism has not been clearly identified. We offer an interpretation of this behavior in Sec. V.

\section{Time domain analysis}
\begin{figure}
\centering\includegraphics[width=8cm]{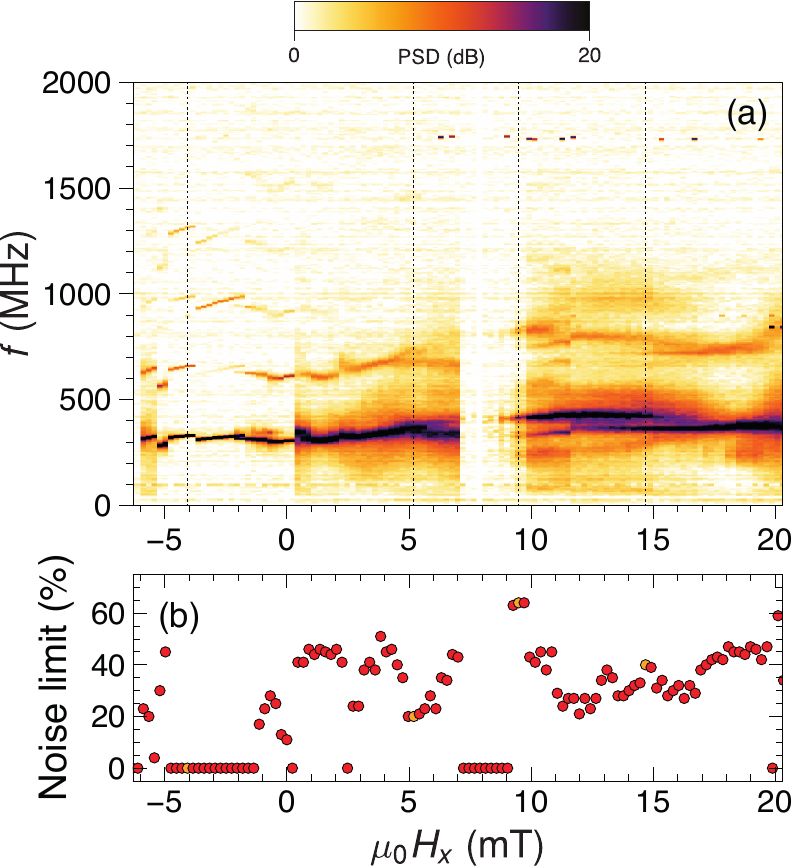}
\caption{(a) Color map of the power spectral density (PSD) computed from time-domain measurements as a function of applied in-plane magnetic field, $\mu_0 H_x$, at a current of 7.5 mA. The dashed lines indicate field values at which time-averaged power spectral densities and spectrograms are shown in Fig.~\ref{fig:PSDspectro}. (b) Noise limit computed with the noise titration technique where nonzero values are consistent with chaotic dynamics. The points in light orange correspond to the field values indicated by the dashed lines in (a).}
\label{fig:timedomainPSDNL}
\end{figure}

To gain further insight into the origin of these multimode frequency spectra, we performed time domain measurements with a single-shot oscilloscope for a range of applied currents and in-plane magnetic fields. In Fig.~\ref{fig:timedomainPSDNL}, we present the PSD as a function of applied magnetic field under a fixed DC current of 7.5 mA.
In Fig.~\ref{fig:timedomainPSDNL}(a), the Welch method \cite{welch_use_1967} was employed to construct the average PSD from the time traces. At each value of the applied field, we recorded a 10-$\mu$s-long time trace which was then decomposed into half-overlapping 100-ns windows. After applying a Hann filter to the time series data to minimize spectral leakage, a discrete Fourier transform was extracted from the windowed data. The power spectrum presented in Fig.~\ref{fig:timedomainPSDNL}(a) represents an average over these windows.

In a similar way to how the spectral characteristics vary under applied perpendicular magnetic fields~\cite{devolder_chaos_2019}, we observe a variety of behaviors as the in-plane applied field is changed at fixed current. We can identify a number of qualitatively distinct features, indicated by the dashed lines, such as narrow spectral lines, line broadening, spectral power decrease, and the presence of several modes. To better understand whether these features, in particular, spectral line broadening, are associated with chaotic dynamics, we use the noise titration technique to quantify whether highly nonlinear dynamics are present in physical processes~\cite{poon_titration_2001, hu_chaos_2006, devolder_chaos_2019}. This technique compares the accuracy of linear and nonlinear autoregressive models based on Volterra-Wiener series of degree $d$ and memory depth $\kappa$ as one-step predictors of the data. Truncated optimal linear and nonlinear models are selected to minimize the Akaike criterion and then statistically tested against each other (using an F test or a Whitney-Mann test) to decide on the best model fitting the data. To detect the presence of chaos, the aforementioned comparison between the two optimal models is performed on the experimental time series corrupted by numerical additive white Gaussian noise with increasing variance. The variance level leading for the first time to a lower normalized residual sum of squares from the linear model is called the noise limit (NL). A sufficient condition for the detection of chaos is then NL$>0$ and its magnitude (expressed in $\%$) represents the strength of chaos, similar to the value of the largest Lyapunov exponent characterizing the average rate of exponential divergence between nearby trajectories in phase space~\cite{kantz_nonlinear_2003}. For our analysis, we have considered a degree $d=2$ with a memory depth $\kappa = 10$ and used a time-delay embedding of $\tau_e = 100 \, \Delta t$ with $\Delta t = 12.5$ ps, the sampling period of the experimental time series. We have also concentrated our analysis on data that was band-pass filtered in the frequency band $[100 \text{ MHz},1 \text{ GHz}]$, where the relevant dynamical features of the nanocontact vortex oscillator are concentrated. The time-delay embedding allows us to sub-sample the time series and remove the sample-to-sample correlations, which would otherwise lead our experimental data to always be better described by linear models \cite{Poon_NLD_1997}. In Fig.~\ref{fig:timedomainPSDNL}(b), we have plotted the NL as a function of the external magnetic field $\mu_0 H_x$ and we can observe that significant, nonzero values of NL correlate with regions in the PSD map in which spectral line broadening appears or where several modes are present. A vanishing NL is observed for negative in-plane fields where a strong spectral line is seen, and in the range of 7 to 10 mT where a loss in the output occurs.

\begin{figure*}
\centering\includegraphics[width=18cm]{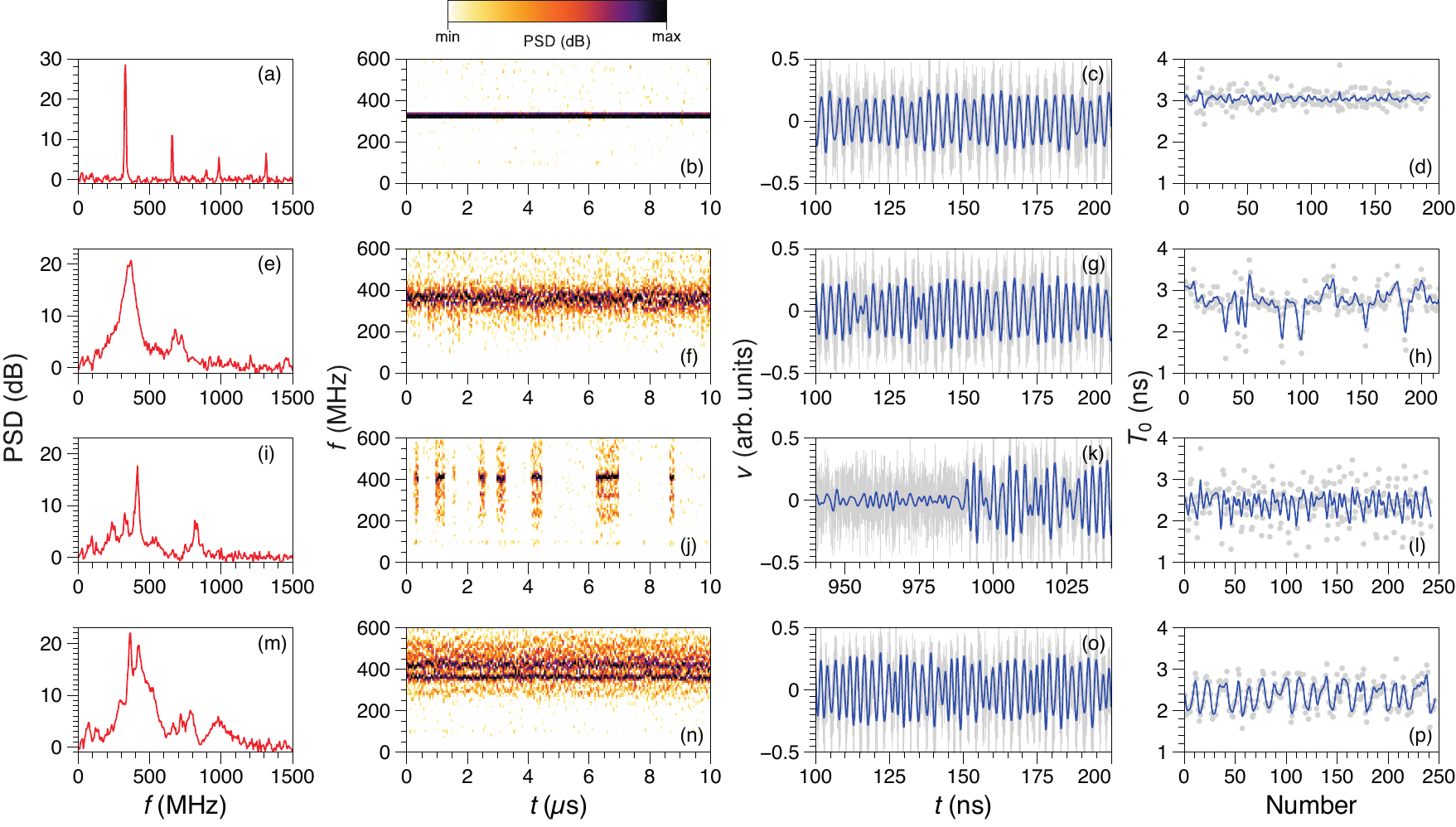}
\caption{Analysis of time series data at 7.5 mA for the four values of the in-plane magnetic field $\mu_0 H_x$ indicated by dashed lines in Fig.~\ref{fig:timedomainPSDNL}(a): (a)--(d) $-4.1$ mT, (e)--(h) 5.2 mT, (i)--(l) 9.5 mT, and (m)--(p) 14.7 mT.  The power spectrum computed with half-overlapping 100-ns windows are shown in (a), (e), (i), (m), with the corresponding spectrograms presented in (b), (f), (j), (n). Examples of raw (gray) and band pass (200--600 MHz) filtered (blue) time traces over an interval of 100 ns are given in (c), (g), (k), (o). The instantaneous period computed from successive zero crossings of the filtered time traces over an interval of 300 ns is given in (d), (h), (l), (p) where both the single (gray dots) and five-point moving averaged values (blue lines) are shown.}
\label{fig:PSDspectro}
\end{figure*}

To shed light on the nature of the different dynamical regimes observed, indicated by the dashed lines in Fig.~\ref{fig:timedomainPSDNL}(a), we provide a deeper analysis of the time series data with a number of different methods as shown in Fig.~\ref{fig:PSDspectro}. 
The case at $\mu_0 H_x = -4.1$ mT at which a single mode excitation is observed [Figs.~\ref{fig:PSDspectro}(a)$-$\ref{fig:PSDspectro}(d)] serves to illustrates these methods.  Figure~\ref{fig:PSDspectro}(a) shows the computed PSD using the Welch method, where we can observe narrow spectral lines consistent with the stable mode gyration, similar to the case shown in Fig.~\ref{heuslerpsd}(a). Figure~\ref{fig:PSDspectro}(b) illustrates the spectrogram constructed from Fourier transforms of the 100-ns half-overlapping windows, which shows that the oscillation mode frequency remains stable over the entire duration of the time trace. In Fig.~\ref{fig:PSDspectro}(c), we give an example of the measured oscillation signal over a 100-ns window, where both the raw and bandpass filtered signal (with a rectangular filter between 200 and 600 MHz) are shown. The filtered signal represents a well-defined oscillatory signal with small fluctuations in the amplitude. To quantify the frequency stability of the mode with a higher temporal resolution than that afforded by the spectrogram, we obtain a crude estimate of the instantaneous frequency by computing the interval between successive zero-crossings of the filtered signal in Fig.~\ref{fig:PSDspectro}(c) and attributing this to a half oscillation period, $T_0/2$. The evolution of $T_0$ with time is given in Fig.~\ref{fig:PSDspectro}(d), where both the raw data and five-point moving average is shown over an interval of 300 ns. This behavior further confirms the stability of this mode, where only small fluctuations about the mean value occur.

Let us now discuss the case of small positive magnetic fields, $\mu_0 H_x = 5.2$ mT, at which broadening of power spectrum can be inferred from Fig.~\ref{fig:timedomainPSDNL}(a). The corresponding power spectrum, shown in Fig.~\ref{fig:PSDspectro}(e), exhibits a broad peak at the fundamental frequency and is reminiscent of the case in Fig.~\ref{heuslerpsd}(c). In contrast to the stable gyration mode [Fig.~\ref{fig:PSDspectro}(b)], the spectrogram for this case in Fig.~\ref{fig:PSDspectro}(f) shows a strong jitter in the mode frequency, which remains localized within a range of $\sim$100 MHz but fluctuates strongly in time. The noise level at this field value is about 40\% [Fig.~\ref{fig:timedomainPSDNL}(b)], which is suggestive of a chaotic process. The time trace in Fig.~\ref{fig:PSDspectro}(g) does not reveal any qualitatively different features as the previous case, but an analysis of the instantaneous period reveals large fluctuations over a range exceeding 1 ns, which appear to be randomly distributed in time [Fig.~\ref{fig:PSDspectro}(h)].

As we noted previously, there is a range of applied fields from approximately 7 to 9 mT in which a microwave-quiet region is observed. As the in-plane field is further increased beyond this range, the reappearance of the microwave signal involves an intermittent signal. An example of this behavior occurs at $\mu_0 H_x = 9.5$ mT for which the power spectrum is shown in Fig.~\ref{fig:PSDspectro}(i). In contrast to the other cases discussed until now, the power spectrum exhibits additional sidebands below the fundamental frequency peak. The spectrogram in Fig.~\ref{fig:PSDspectro}(j) shows a largely quiet spectrum that is punctuated by intermittent oscillations with the peak structure in Fig.~\ref{fig:PSDspectro}(i). While statistics on such events are limited, the intermittence appears to be aperiodic and is not accompanied by any discernible transient dynamics –  at least within the time scale given by the 100-ns window used to compute the spectrogram. This can be seen in Fig.~\ref{fig:PSDspectro}(k) where the oscillations emerge from the noise (at the center of the time window) without any additional transients.  When present, the oscillatory signal exhibits additional oscillations in the amplitude, leading to the sidebands below the fundamental frequency, which is suggestive of a beat pattern. This is further corroborated by the analysis of the instantaneous period in Fig.~\ref{fig:PSDspectro}(l), where a regular variation over a range of $\approx 0.8$ ns can be seen. The proportion of the intermittent signal that occupies the entire 10-$\mu$s window studied increases as the field is increased; there is 0\% of signal at 7.9 mT, 7\% at 8.8 mT, 22\% at 9.5 mT, 81\% at 10.2 mT, and 100\% at 13.8 mT.

Finally, under a higher in-plane applied field of $\mu_0 H_x = 14.7$ mT, we observe a double-peak structure with a broad base [Fig.~\ref{fig:PSDspectro}(m)], with some discernible modulation side bands that are a few dB above the noise floor. The spectrogram in Fig.~\ref{fig:PSDspectro}(n) reveals that spectral power remains present at both frequency branches of $\sim 375$ MHz and $\sim 425$ MHz across the entire time trace,  which is suggestive of mode coexistence rather than mode hopping.  In the time trace shown in Fig.~\ref{fig:PSDspectro}(o), we can discern some variations in the period of the filtered signal. This is quantified in Fig.~\ref{fig:PSDspectro}(p) where oscillations in the instantaneous period can be seen. Like the example in Fig.~\ref{fig:PSDspectro}(l), this behavior is reminiscent of frequency modulation in which the oscillator frequency varies periodically in time.

\begin{figure}
\centering\includegraphics[width=8.5cm]{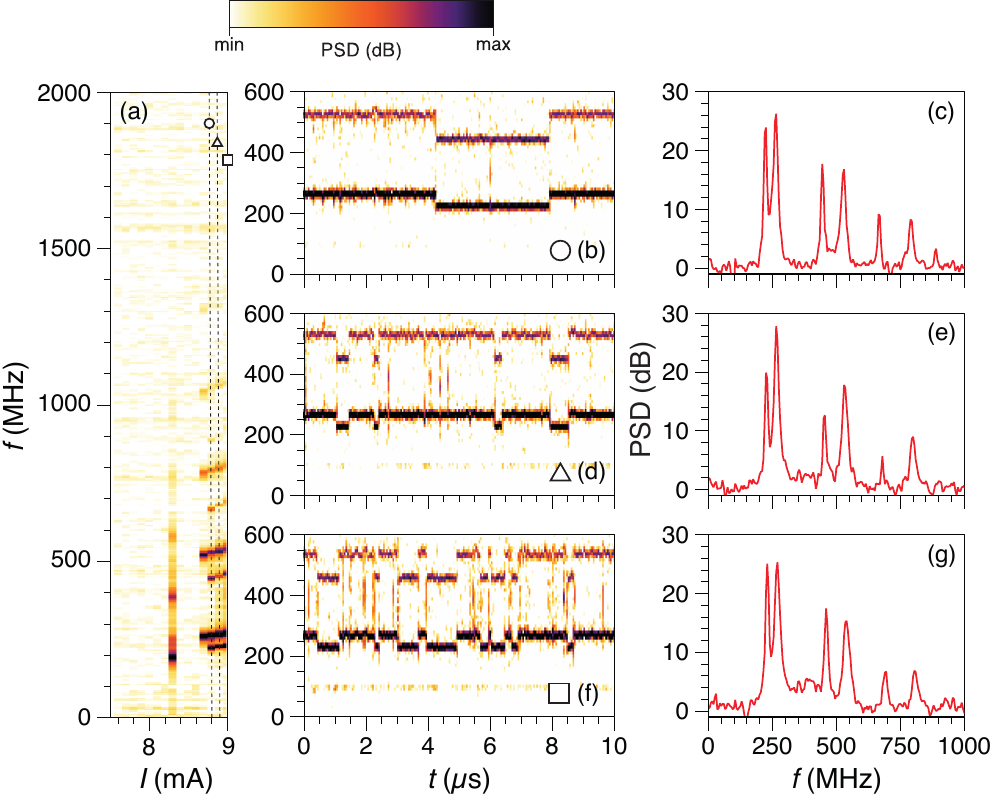}
\caption{(a) Power spectral density computed from time-domain measurements as a function of applied current, $I$, under an applied field of $\mu_0 H_x = -0.7$ mT. The dashed lines labeled with symbols indicate current values at which spectrograms constructed with half-overlapping 100-ns windows (b), (d), (f) and the corresponding PSD computed with the Welch method are shown: (b), (c) 8.8 mA, (d),(e) 8.9 mA, and (f), (g) 9.0 mA.}
\label{fig:varI_timedomain}
\end{figure}

Clear evidence of mode hopping was observed in another set of experiments in which the applied current was varied in a small interval, as shown in Fig.~\ref{fig:varI_timedomain}(a).
Here, the range of applied currents in which microwave power is observed is less than 1 mA, where for decreasing currents a double-mode spectrum can be seen followed by a microwave-quiet region with the exception at 8.3 mA, which is reminiscent of the case in Fig.~\ref{heuslerpsd}(e). The spectrograms for three values of the applied current are shown in Figs.~\ref{fig:varI_timedomain}(b), \ref{fig:varI_timedomain}(d), and \ref{fig:varI_timedomain}(f), where clear transitions can be seen between two well-defined modes. While the statistics are limited, the dwell times appear to fluctuate randomly between the two states, with transitions being more frequent as the current is increased. The corresponding power spectra are shown in Figs.~\ref{fig:varI_timedomain}(c), \ref{fig:varI_timedomain}(e), and \ref{fig:varI_timedomain}(g), which exhibit a clear double-peak structure that more closely resembles Fig.~\ref{heuslercurves}(d) than Fig.~\ref{fig:PSDspectro}(m).

\section{Discussion and concluding remarks \label{conclu}}

Clear signatures of vortex gyration were detected in a similar current and frequency range to behavior previously reported in permalloy-based nanocontact systems~\cite{petit-watelot_commensurability_2012, devolder_chaos_2019, letang_modulation_2019}. There are, however, a few notable differences. First, core reversal does not appear to be triggered in the Heusler-based systems studied here, despite similar gyration frequencies being reached at the upper limit of the applied current. The onset of periodic core reversal results in the appearance of modulation sidebands, a feature that has not been observed in the Heusler-based nanocontact discussed here. This absence may be due to the fact that sufficiently large current densities have not been reached, since the onset of core reversal typically occurs above 10 mA in permalloy-based systems. Another possibility is that the larger exchange constant (Co$_2$MnGe value being half to full Co value, i.e., 15 to 30 pJ/m, which is higher than permalloy or Ni value, which is 10 pJ/m) means that the critical deformation is not reached for core reversal under the spin torques applied~\cite{tretiakov_vortices_2007, guslienko_dynamic_2008, yoo_analytical_2021}. Alternatively, core reversal is, in fact, present, but the low damping values result in a long relaxation time between each core reversal event. As a result, the modulation frequency is comparatively low and the sideband is indiscernible from the main oscillation peak.

Mode hopping~\cite{pribiag_long-timescale_2009, kuepferling_two_2010, muduli_decoherence_2012, Heinonen:2013iu, Dumas:2013bx, keatley_imaging_2017, zhang_tunable_2017}, mode coexistence \cite{krivorotov_large-amplitude_2007, wang_multiple-mode_2011, Iacocca:2015ev, zhang_tunable_2017}, and mode variations with \idc~\cite{wang_three-mode_2011} have already been reported in the literature for other oscillator systems. Mode coexistence can be attributed to two oscillation modes, for example, where an oscillatory mode is present in each of the magnetic layers of a coupled bilayer system. On the other hand, mode hopping describes the transition, quite likely thermally activated, between two distinct oscillatory regimes within a given system, which possess distinct frequencies. Our observations in the time domain appear to be consistent with both of these scenarios, where the variation of the in-plane field allows us to transit from one behavior to the other.

Preliminary micromagnetics simulations of the nucleation process indicate that a vortex can be nucleated in both the free and reference layers. While dipolar coupling between the layers is minimized in the uniformly in-plane magnetized state, it can become significant when vortices are present in both layers. In nanopillar structures, dipolar coupling between vortices in the free and reference layers can result in coupled dynamics~\cite{Locatelli:2011hw, lebrun_nonlinear_2014}, which leads to narrow spectral lines. Here, we can offer an interpretation of the observed behavior in Figs.~\ref{fig:timedomainPSDNL}--\ref{fig:varI_timedomain} in terms of coupled vortex dynamics, where the vortices in the free and reference layers both undergo steady-state gyrations as a result of the CIP torques. A single stable frequency signal, such as in Fig.~\ref{fig:PSDspectro}(a)--\ref{fig:PSDspectro}(d), results from the synchronized motion of the two vortices, leading to improved phase coherence and narrower spectral lines. Conditions for synchronization degrade as the in-plane field is increased, leading to unsynchronized motion along independent trajectories with close frequencies, as shown in Figs.~\ref{fig:PSDspectro}(m) and \ref{fig:PSDspectro}(n) [and possibly also in Figs.~\ref{fig:PSDspectro}(e) and \ref{fig:PSDspectro}(f)]. Through the dipolar coupling, the resulting dynamics shows signs of chaos as detected from noise titration, which is not unexpected for coupled oscillators. The intermittent signal in Figs.~\ref{fig:PSDspectro}(i) and \ref{fig:PSDspectro}(j) does not exhibit any signs of transient dynamics, for example, associated with oscillator death, so we can interpret the loss in microwave power as resulting from a synchronized regime in which the oscillators are in anti-phase, which leads to a drastic reduction in the overall time-varying MR of the pseudo spin valve. As the in-plane field is varied outside of this range of intermittence, we observe a clearer separation between the frequency of the two gyration modes, which is particularly evident in Figs.~\ref{fig:PSDspectro}(m) and \ref{fig:PSDspectro}(n) with the presence of two distinct branches. In this light, we can then interpret the mode hopping in Figs.~\ref{fig:varI_timedomain}(b), \ref{fig:varI_timedomain}(d), and \ref{fig:varI_timedomain}(f) as synchronized motion in which the master/slave relationship alternates between the free and reference layer oscillators. This is consistent with the narrower spectral lines seen as compared with the two-mode behavior in Figs.~\ref{fig:PSDspectro}(m) and \ref{fig:PSDspectro}(n).

Full micromagnetics simulations of the coupled bilayer system prove to be particularly challenging to implement primarily because of the low values of the Gilbert damping constant. In systems with self-sustained oscillations, the damping constant enters as a prefactor in the relaxation rate of amplitude fluctuations toward the limit cycle~\cite{kim_spin-torque_2012}. Because the value of $\alpha$ in the Co$_2$MnGe films studied is about 20 times smaller than its counterpart in permalloy ($\alpha \sim 0.01$), it means that transient dynamics associated with changes in applied currents or fields, which might last 25 ns in a permalloy-based system, would take up to 500 ns to damp out in the Heusler system. In order words, the transient dynamics toward the limit cycle would take much longer to simulate than several periods of gyration itself, which is needed to compute simulated maps of the PSD. Further simulations will be pursued but a detailed numerical study remains beyond the scope of the present paper.

Finally, we note that the signal-to-noise ratio of the pseudo spin valve is not significantly higher than that of permalloy-based nanocontact vortex oscillators we studied previously. This is unexpected since we anticipated a strong MR signal based on the higher spin polarization of the Heusler compound used. We speculate that this is a consequence of the nanocontact geometry, where the large current densities flow perpendicular to the film plane and also within the film plane~\cite{petit-watelot_electrical_2012}. The overall electrical signal is therefore based on a combination of CPP-GMR and CIP-GMR, with possible contributions also from AMR related to the CIP flow through the vortex structure. As such, the half-metallic nature of the Heusler compound is not fully exploited by the device geometry, which is not the case, for example, for oscillators based on nanopillars.

In summary, we present an experimental study of the current-driven magnetization dynamics in nanocontact oscillators based on epitaxially-grown pseudo spin valve structures with the Heusler compound Co$_2$MnGe. The hard ferromagnetic layer was successfully fabricated as a bottom electrode by growing the Co$_2$MnGe film biased through exchange coupling to an epitaxial Co layer. The absence of interlayer RKKY coupling between the free and reference layers was shown for spacer thicknesses of 2 nm with low values of the Gilbert damping constant in the range of $6 \times 10^{-4}$ preserved at room temperature. We have demonstrated that vortex dynamics in nanocontact oscillators based on Heusler compounds can lead to rich power spectra, which comprise phenomena such as mode coexistence and mode hopping. Such features are possibly linked to the dipolar-coupled magnetization dynamics of the vortex cores in the free and reference layers, and accentuated by the low values of the Gilbert damping constant.

\begin{acknowledgments}
The work was supported by the Agence Nationale de la Recherche under Contract No. ANR-17-CE24-0008 (CHIPMuNCS). The authors thank S. Migot for FIB preparation and J. Ghanbaja for TEM imaging. J.L. acknowledges financial support from the EOBE Doctoral School of Universit{\'e} Paris-Saclay. D.R. gratefully thanks C.-S. Poon for sharing information on the implementation of the titration technique. D.R. acknowledges  financial support from the Chair in Photonics.
\end{acknowledgments}

\bibliography{main.bbl} 

\begin{thebibliography}{69}%
\makeatletter
\providecommand \@ifxundefined [1]{%
 \@ifx{#1\undefined}
}%
\providecommand \@ifnum [1]{%
 \ifnum #1\expandafter \@firstoftwo
 \else \expandafter \@secondoftwo
 \fi
}%
\providecommand \@ifx [1]{%
 \ifx #1\expandafter \@firstoftwo
 \else \expandafter \@secondoftwo
 \fi
}%
\providecommand \natexlab [1]{#1}%
\providecommand \enquote  [1]{``#1''}%
\providecommand \bibnamefont  [1]{#1}%
\providecommand \bibfnamefont [1]{#1}%
\providecommand \citenamefont [1]{#1}%
\providecommand \href@noop [0]{\@secondoftwo}%
\providecommand \href [0]{\begingroup \@sanitize@url \@href}%
\providecommand \@href[1]{\@@startlink{#1}\@@href}%
\providecommand \@@href[1]{\endgroup#1\@@endlink}%
\providecommand \@sanitize@url [0]{\catcode `\\12\catcode `\$12\catcode
  `\&12\catcode `\#12\catcode `\^12\catcode `\_12\catcode `\%12\relax}%
\providecommand \@@startlink[1]{}%
\providecommand \@@endlink[0]{}%
\providecommand \url  [0]{\begingroup\@sanitize@url \@url }%
\providecommand \@url [1]{\endgroup\@href {#1}{\urlprefix }}%
\providecommand \urlprefix  [0]{URL }%
\providecommand \Eprint [0]{\href }%
\providecommand \doibase [0]{https://doi.org/}%
\providecommand \selectlanguage [0]{\@gobble}%
\providecommand \bibinfo  [0]{\@secondoftwo}%
\providecommand \bibfield  [0]{\@secondoftwo}%
\providecommand \translation [1]{[#1]}%
\providecommand \BibitemOpen [0]{}%
\providecommand \bibitemStop [0]{}%
\providecommand \bibitemNoStop [0]{.\EOS\space}%
\providecommand \EOS [0]{\spacefactor3000\relax}%
\providecommand \BibitemShut  [1]{\csname bibitem#1\endcsname}%
\let\auto@bib@innerbib\@empty
\bibitem [{\citenamefont {Choi}\ \emph {et~al.}(2014)\citenamefont {Choi},
  \citenamefont {Kang}, \citenamefont {Cho}, \citenamefont {Oh}, \citenamefont
  {Shin}, \citenamefont {Park}, \citenamefont {Jang}, \citenamefont {Min},
  \citenamefont {Kim}, \citenamefont {Park},\ and\ \citenamefont
  {Park}}]{choi_spin_2014}%
  \BibitemOpen
  \bibfield  {author} {\bibinfo {author} {\bibfnamefont {H.~S.}\ \bibnamefont
  {Choi}}, \bibinfo {author} {\bibfnamefont {S.~Y.}\ \bibnamefont {Kang}},
  \bibinfo {author} {\bibfnamefont {S.~J.}\ \bibnamefont {Cho}}, \bibinfo
  {author} {\bibfnamefont {I.-Y.}\ \bibnamefont {Oh}}, \bibinfo {author}
  {\bibfnamefont {M.}~\bibnamefont {Shin}}, \bibinfo {author} {\bibfnamefont
  {H.}~\bibnamefont {Park}}, \bibinfo {author} {\bibfnamefont {C.}~\bibnamefont
  {Jang}}, \bibinfo {author} {\bibfnamefont {B.-C.}\ \bibnamefont {Min}},
  \bibinfo {author} {\bibfnamefont {S.-I.}\ \bibnamefont {Kim}}, \bibinfo
  {author} {\bibfnamefont {S.-Y.}\ \bibnamefont {Park}},\ and\ \bibinfo
  {author} {\bibfnamefont {C.~S.}\ \bibnamefont {Park}},\ }\bibfield  {title}
  {\bibinfo {title} {Spin nano{\textendash}oscillator{\textendash}based
  wireless communication},\ }\href {https://doi.org/10.1038/srep05486}
  {\bibfield  {journal} {\bibinfo  {journal} {Scientific Reports}\ }\textbf
  {\bibinfo {volume} {4}},\ \bibinfo {pages} {5486} (\bibinfo {year}
  {2014})}\BibitemShut {NoStop}%
\bibitem [{\citenamefont {Ruiz-Calaforra}\ \emph {et~al.}(2017)\citenamefont
  {Ruiz-Calaforra}, \citenamefont {Purbawati}, \citenamefont {Br{\"a}cher},
  \citenamefont {Hem}, \citenamefont {Murapaka}, \citenamefont {Jim{\'e}nez},
  \citenamefont {Mauri}, \citenamefont {Zeltser}, \citenamefont {Katine},
  \citenamefont {Cyrille}, \citenamefont {Buda-Prejbeanu},\ and\ \citenamefont
  {Ebels}}]{ruiz-calaforra_frequency_2017}%
  \BibitemOpen
  \bibfield  {author} {\bibinfo {author} {\bibfnamefont {A.}~\bibnamefont
  {Ruiz-Calaforra}}, \bibinfo {author} {\bibfnamefont {A.}~\bibnamefont
  {Purbawati}}, \bibinfo {author} {\bibfnamefont {T.}~\bibnamefont
  {Br{\"a}cher}}, \bibinfo {author} {\bibfnamefont {J.}~\bibnamefont {Hem}},
  \bibinfo {author} {\bibfnamefont {C.}~\bibnamefont {Murapaka}}, \bibinfo
  {author} {\bibfnamefont {E.}~\bibnamefont {Jim{\'e}nez}}, \bibinfo {author}
  {\bibfnamefont {D.}~\bibnamefont {Mauri}}, \bibinfo {author} {\bibfnamefont
  {A.}~\bibnamefont {Zeltser}}, \bibinfo {author} {\bibfnamefont {J.~A.}\
  \bibnamefont {Katine}}, \bibinfo {author} {\bibfnamefont {M.-C.}\
  \bibnamefont {Cyrille}}, \bibinfo {author} {\bibfnamefont {L.~D.}\
  \bibnamefont {Buda-Prejbeanu}},\ and\ \bibinfo {author} {\bibfnamefont
  {U.}~\bibnamefont {Ebels}},\ }\bibfield  {title} {\bibinfo {title} {Frequency
  shift keying by current modulation in a {MTJ}-based {STNO} with high data
  rate},\ }\href {https://doi.org/10.1063/1.4994892} {\bibfield  {journal}
  {\bibinfo  {journal} {Applied Physics Letters}\ }\textbf {\bibinfo {volume}
  {111}},\ \bibinfo {pages} {082401} (\bibinfo {year} {2017})}\BibitemShut
  {NoStop}%
\bibitem [{\citenamefont {Tulapurkar}\ \emph {et~al.}(2005)\citenamefont
  {Tulapurkar}, \citenamefont {Suzuki}, \citenamefont {Fukushima},
  \citenamefont {Kubota}, \citenamefont {Maehara}, \citenamefont {Tsunekawa},
  \citenamefont {Djayaprawira}, \citenamefont {Watanabe},\ and\ \citenamefont
  {Yuasa}}]{tulapurkar_spin-torque_2005}%
  \BibitemOpen
  \bibfield  {author} {\bibinfo {author} {\bibfnamefont {A.~A.}\ \bibnamefont
  {Tulapurkar}}, \bibinfo {author} {\bibfnamefont {Y.}~\bibnamefont {Suzuki}},
  \bibinfo {author} {\bibfnamefont {A.}~\bibnamefont {Fukushima}}, \bibinfo
  {author} {\bibfnamefont {H.}~\bibnamefont {Kubota}}, \bibinfo {author}
  {\bibfnamefont {H.}~\bibnamefont {Maehara}}, \bibinfo {author} {\bibfnamefont
  {K.}~\bibnamefont {Tsunekawa}}, \bibinfo {author} {\bibfnamefont {D.~D.}\
  \bibnamefont {Djayaprawira}}, \bibinfo {author} {\bibfnamefont
  {N.}~\bibnamefont {Watanabe}},\ and\ \bibinfo {author} {\bibfnamefont
  {S.}~\bibnamefont {Yuasa}},\ }\bibfield  {title} {\bibinfo {title}
  {Spin-torque diode effect in magnetic tunnel junctions},\ }\href
  {https://doi.org/10.1038/nature04207} {\bibfield  {journal} {\bibinfo
  {journal} {Nature}\ }\textbf {\bibinfo {volume} {438}},\ \bibinfo {pages}
  {339} (\bibinfo {year} {2005})}\BibitemShut {NoStop}%
\bibitem [{\citenamefont {Jenkins}\ \emph {et~al.}(2016)\citenamefont
  {Jenkins}, \citenamefont {Lebrun}, \citenamefont {Grimaldi}, \citenamefont
  {Tsunegi}, \citenamefont {Bortolotti}, \citenamefont {Kubota}, \citenamefont
  {Yakushiji}, \citenamefont {Fukushima}, \citenamefont {de~Loubens},
  \citenamefont {Klein}, \citenamefont {Yuasa},\ and\ \citenamefont
  {Cros}}]{jenkins_spin-torque_2016}%
  \BibitemOpen
  \bibfield  {author} {\bibinfo {author} {\bibfnamefont {A.~S.}\ \bibnamefont
  {Jenkins}}, \bibinfo {author} {\bibfnamefont {R.}~\bibnamefont {Lebrun}},
  \bibinfo {author} {\bibfnamefont {E.}~\bibnamefont {Grimaldi}}, \bibinfo
  {author} {\bibfnamefont {S.}~\bibnamefont {Tsunegi}}, \bibinfo {author}
  {\bibfnamefont {P.}~\bibnamefont {Bortolotti}}, \bibinfo {author}
  {\bibfnamefont {H.}~\bibnamefont {Kubota}}, \bibinfo {author} {\bibfnamefont
  {K.}~\bibnamefont {Yakushiji}}, \bibinfo {author} {\bibfnamefont
  {A.}~\bibnamefont {Fukushima}}, \bibinfo {author} {\bibfnamefont
  {G.}~\bibnamefont {de~Loubens}}, \bibinfo {author} {\bibfnamefont
  {O.}~\bibnamefont {Klein}}, \bibinfo {author} {\bibfnamefont
  {S.}~\bibnamefont {Yuasa}},\ and\ \bibinfo {author} {\bibfnamefont
  {V.}~\bibnamefont {Cros}},\ }\bibfield  {title} {\bibinfo {title}
  {Spin-torque resonant expulsion of the vortex core for an efficient
  radiofrequency detection scheme},\ }\href
  {https://doi.org/10.1038/nnano.2015.295} {\bibfield  {journal} {\bibinfo
  {journal} {Nature Nanotechnology}\ }\textbf {\bibinfo {volume} {11}},\
  \bibinfo {pages} {360} (\bibinfo {year} {2016})}\BibitemShut {NoStop}%
\bibitem [{\citenamefont {Torrejon}\ \emph {et~al.}(2017)\citenamefont
  {Torrejon}, \citenamefont {Riou}, \citenamefont {Araujo}, \citenamefont
  {Tsunegi}, \citenamefont {Khalsa}, \citenamefont {Querlioz}, \citenamefont
  {Bortolotti}, \citenamefont {Cros}, \citenamefont {Yakushiji}, \citenamefont
  {Fukushima}, \citenamefont {Kubota}, \citenamefont {Yuasa}, \citenamefont
  {Stiles},\ and\ \citenamefont {Grollier}}]{torrejon_neuromorphic_2017}%
  \BibitemOpen
  \bibfield  {author} {\bibinfo {author} {\bibfnamefont {J.}~\bibnamefont
  {Torrejon}}, \bibinfo {author} {\bibfnamefont {M.}~\bibnamefont {Riou}},
  \bibinfo {author} {\bibfnamefont {F.~A.}\ \bibnamefont {Araujo}}, \bibinfo
  {author} {\bibfnamefont {S.}~\bibnamefont {Tsunegi}}, \bibinfo {author}
  {\bibfnamefont {G.}~\bibnamefont {Khalsa}}, \bibinfo {author} {\bibfnamefont
  {D.}~\bibnamefont {Querlioz}}, \bibinfo {author} {\bibfnamefont
  {P.}~\bibnamefont {Bortolotti}}, \bibinfo {author} {\bibfnamefont
  {V.}~\bibnamefont {Cros}}, \bibinfo {author} {\bibfnamefont {K.}~\bibnamefont
  {Yakushiji}}, \bibinfo {author} {\bibfnamefont {A.}~\bibnamefont
  {Fukushima}}, \bibinfo {author} {\bibfnamefont {H.}~\bibnamefont {Kubota}},
  \bibinfo {author} {\bibfnamefont {S.}~\bibnamefont {Yuasa}}, \bibinfo
  {author} {\bibfnamefont {M.~D.}\ \bibnamefont {Stiles}},\ and\ \bibinfo
  {author} {\bibfnamefont {J.}~\bibnamefont {Grollier}},\ }\bibfield  {title}
  {\bibinfo {title} {Neuromorphic computing with nanoscale spintronic
  oscillators},\ }\href {https://doi.org/10.1038/nature23011} {\bibfield
  {journal} {\bibinfo  {journal} {Nature}\ }\textbf {\bibinfo {volume} {547}},\
  \bibinfo {pages} {428} (\bibinfo {year} {2017})}\BibitemShut {NoStop}%
\bibitem [{\citenamefont {Romera}\ \emph {et~al.}(2018)\citenamefont {Romera},
  \citenamefont {Talatchian}, \citenamefont {Tsunegi}, \citenamefont {Araujo},
  \citenamefont {Cros}, \citenamefont {Bortolotti}, \citenamefont {Trastoy},
  \citenamefont {Yakushiji}, \citenamefont {Fukushima}, \citenamefont {Kubota},
  \citenamefont {Yuasa}, \citenamefont {Ernoult}, \citenamefont
  {Vodenicarevic}, \citenamefont {Hirtzlin}, \citenamefont {Locatelli},
  \citenamefont {Querlioz},\ and\ \citenamefont
  {Grollier}}]{romera_vowel_2018}%
  \BibitemOpen
  \bibfield  {author} {\bibinfo {author} {\bibfnamefont {M.}~\bibnamefont
  {Romera}}, \bibinfo {author} {\bibfnamefont {P.}~\bibnamefont {Talatchian}},
  \bibinfo {author} {\bibfnamefont {S.}~\bibnamefont {Tsunegi}}, \bibinfo
  {author} {\bibfnamefont {F.~A.}\ \bibnamefont {Araujo}}, \bibinfo {author}
  {\bibfnamefont {V.}~\bibnamefont {Cros}}, \bibinfo {author} {\bibfnamefont
  {P.}~\bibnamefont {Bortolotti}}, \bibinfo {author} {\bibfnamefont
  {J.}~\bibnamefont {Trastoy}}, \bibinfo {author} {\bibfnamefont
  {K.}~\bibnamefont {Yakushiji}}, \bibinfo {author} {\bibfnamefont
  {A.}~\bibnamefont {Fukushima}}, \bibinfo {author} {\bibfnamefont
  {H.}~\bibnamefont {Kubota}}, \bibinfo {author} {\bibfnamefont
  {S.}~\bibnamefont {Yuasa}}, \bibinfo {author} {\bibfnamefont
  {M.}~\bibnamefont {Ernoult}}, \bibinfo {author} {\bibfnamefont
  {D.}~\bibnamefont {Vodenicarevic}}, \bibinfo {author} {\bibfnamefont
  {T.}~\bibnamefont {Hirtzlin}}, \bibinfo {author} {\bibfnamefont
  {N.}~\bibnamefont {Locatelli}}, \bibinfo {author} {\bibfnamefont
  {D.}~\bibnamefont {Querlioz}},\ and\ \bibinfo {author} {\bibfnamefont
  {J.}~\bibnamefont {Grollier}},\ }\bibfield  {title} {\bibinfo {title} {Vowel
  recognition with four coupled spin-torque nano-oscillators},\ }\href
  {https://doi.org/10.1038/s41586-018-0632-y} {\bibfield  {journal} {\bibinfo
  {journal} {Nature}\ }\textbf {\bibinfo {volume} {563}},\ \bibinfo {pages}
  {230} (\bibinfo {year} {2018})}\BibitemShut {NoStop}%
\bibitem [{\citenamefont {Tsunegi}\ \emph {et~al.}(2019)\citenamefont
  {Tsunegi}, \citenamefont {Taniguchi}, \citenamefont {Nakajima}, \citenamefont
  {Miwa}, \citenamefont {Yakushiji}, \citenamefont {Fukushima}, \citenamefont
  {Yuasa},\ and\ \citenamefont {Kubota}}]{tsunegi_physical_2019}%
  \BibitemOpen
  \bibfield  {author} {\bibinfo {author} {\bibfnamefont {S.}~\bibnamefont
  {Tsunegi}}, \bibinfo {author} {\bibfnamefont {T.}~\bibnamefont {Taniguchi}},
  \bibinfo {author} {\bibfnamefont {K.}~\bibnamefont {Nakajima}}, \bibinfo
  {author} {\bibfnamefont {S.}~\bibnamefont {Miwa}}, \bibinfo {author}
  {\bibfnamefont {K.}~\bibnamefont {Yakushiji}}, \bibinfo {author}
  {\bibfnamefont {A.}~\bibnamefont {Fukushima}}, \bibinfo {author}
  {\bibfnamefont {S.}~\bibnamefont {Yuasa}},\ and\ \bibinfo {author}
  {\bibfnamefont {H.}~\bibnamefont {Kubota}},\ }\bibfield  {title} {\bibinfo
  {title} {Physical reservoir computing based on spin torque oscillator with
  forced synchronization},\ }\href {https://doi.org/10.1063/1.5081797}
  {\bibfield  {journal} {\bibinfo  {journal} {Applied Physics Letters}\
  }\textbf {\bibinfo {volume} {114}},\ \bibinfo {pages} {164101} (\bibinfo
  {year} {2019})}\BibitemShut {NoStop}%
\bibitem [{\citenamefont {Williame}\ \emph {et~al.}(2019)\citenamefont
  {Williame}, \citenamefont {Difini~Accioly}, \citenamefont {Rontani},
  \citenamefont {Sciamanna},\ and\ \citenamefont
  {Kim}}]{williame_chaotic_2019}%
  \BibitemOpen
  \bibfield  {author} {\bibinfo {author} {\bibfnamefont {J.}~\bibnamefont
  {Williame}}, \bibinfo {author} {\bibfnamefont {A.}~\bibnamefont
  {Difini~Accioly}}, \bibinfo {author} {\bibfnamefont {D.}~\bibnamefont
  {Rontani}}, \bibinfo {author} {\bibfnamefont {M.}~\bibnamefont {Sciamanna}},\
  and\ \bibinfo {author} {\bibfnamefont {J.-V.}\ \bibnamefont {Kim}},\
  }\bibfield  {title} {\bibinfo {title} {Chaotic dynamics in a macrospin
  spin-torque nano-oscillator with delayed feedback},\ }\href
  {https://doi.org/10.1063/1.5095630} {\bibfield  {journal} {\bibinfo
  {journal} {Applied Physics Letters}\ }\textbf {\bibinfo {volume} {114}},\
  \bibinfo {pages} {232405} (\bibinfo {year} {2019})}\BibitemShut {NoStop}%
\bibitem [{\citenamefont {{Heusler,
  Friedrich}}(1903)}]{heusler_friedrich_uber_1903}%
  \BibitemOpen
  \bibfield  {author} {\bibinfo {author} {\bibnamefont {{Heusler,
  Friedrich}}},\ }\bibfield  {title} {\bibinfo {title} {{\"U}ber magnetische
  {Manganlegierungen}},\ }\href
  {http://archive.org/details/verhandlungende33unkngoog} {\bibfield  {journal}
  {\bibinfo  {journal} {Verhandlungen der Deutschen physikalischen
  Gesellschaft}\ }\textbf {\bibinfo {volume} {12}},\ \bibinfo {pages} {219}
  (\bibinfo {year} {1903})}\BibitemShut {NoStop}%
\bibitem [{\citenamefont {Graf}\ \emph {et~al.}(2011)\citenamefont {Graf},
  \citenamefont {Felser},\ and\ \citenamefont {Parkin}}]{graf_simple_2011}%
  \BibitemOpen
  \bibfield  {author} {\bibinfo {author} {\bibfnamefont {T.}~\bibnamefont
  {Graf}}, \bibinfo {author} {\bibfnamefont {C.}~\bibnamefont {Felser}},\ and\
  \bibinfo {author} {\bibfnamefont {S.~S.~P.}\ \bibnamefont {Parkin}},\
  }\bibfield  {title} {\bibinfo {title} {Simple rules for the understanding of
  {Heusler} compounds},\ }\href
  {https://doi.org/10.1016/j.progsolidstchem.2011.02.001} {\bibfield  {journal}
  {\bibinfo  {journal} {Progress in Solid State Chemistry}\ }\textbf {\bibinfo
  {volume} {39}},\ \bibinfo {pages} {1} (\bibinfo {year} {2011})}\BibitemShut
  {NoStop}%
\bibitem [{\citenamefont {Andrieu}\ \emph {et~al.}(2016)\citenamefont
  {Andrieu}, \citenamefont {Neggache}, \citenamefont {Hauet}, \citenamefont
  {Devolder}, \citenamefont {Hallal}, \citenamefont {Chshiev}, \citenamefont
  {Bataille}, \citenamefont {Le~F{\`e}vre},\ and\ \citenamefont
  {Bertran}}]{andrieu_direct_2016}%
  \BibitemOpen
  \bibfield  {author} {\bibinfo {author} {\bibfnamefont {S.}~\bibnamefont
  {Andrieu}}, \bibinfo {author} {\bibfnamefont {A.}~\bibnamefont {Neggache}},
  \bibinfo {author} {\bibfnamefont {T.}~\bibnamefont {Hauet}}, \bibinfo
  {author} {\bibfnamefont {T.}~\bibnamefont {Devolder}}, \bibinfo {author}
  {\bibfnamefont {A.}~\bibnamefont {Hallal}}, \bibinfo {author} {\bibfnamefont
  {M.}~\bibnamefont {Chshiev}}, \bibinfo {author} {\bibfnamefont {A.~M.}\
  \bibnamefont {Bataille}}, \bibinfo {author} {\bibfnamefont {P.}~\bibnamefont
  {Le~F{\`e}vre}},\ and\ \bibinfo {author} {\bibfnamefont {F.}~\bibnamefont
  {Bertran}},\ }\bibfield  {title} {\bibinfo {title} {Direct evidence for
  minority spin gap in the {Co}${_2}${Mn}{Si} {Heusler} compound},\ }\href
  {https://doi.org/10.1103/PhysRevB.93.094417} {\bibfield  {journal} {\bibinfo
  {journal} {Physical Review B}\ }\textbf {\bibinfo {volume} {93}},\ \bibinfo
  {pages} {094417} (\bibinfo {year} {2016})}\BibitemShut {NoStop}%
\bibitem [{\citenamefont {K{\"u}bler}\ \emph {et~al.}(2007)\citenamefont
  {K{\"u}bler}, \citenamefont {Fecher},\ and\ \citenamefont
  {Felser}}]{kubler_understanding_2007}%
  \BibitemOpen
  \bibfield  {author} {\bibinfo {author} {\bibfnamefont {J.}~\bibnamefont
  {K{\"u}bler}}, \bibinfo {author} {\bibfnamefont {G.~H.}\ \bibnamefont
  {Fecher}},\ and\ \bibinfo {author} {\bibfnamefont {C.}~\bibnamefont
  {Felser}},\ }\bibfield  {title} {\bibinfo {title} {Understanding the trend in
  the {Curie} temperatures of {Co}${_2}$-based {Heusler} compounds: {Ab} initio
  calculations},\ }\href {https://doi.org/10.1103/PhysRevB.76.024414}
  {\bibfield  {journal} {\bibinfo  {journal} {Physical Review B}\ }\textbf
  {\bibinfo {volume} {76}},\ \bibinfo {pages} {024414} (\bibinfo {year}
  {2007})}\BibitemShut {NoStop}%
\bibitem [{\citenamefont {Yamamoto}\ \emph
  {et~al.}(2016{\natexlab{a}})\citenamefont {Yamamoto}, \citenamefont {Seki},
  \citenamefont {Kotsugi},\ and\ \citenamefont
  {Takanashi}}]{yamamoto_magnetic_2016}%
  \BibitemOpen
  \bibfield  {author} {\bibinfo {author} {\bibfnamefont {T.}~\bibnamefont
  {Yamamoto}}, \bibinfo {author} {\bibfnamefont {T.}~\bibnamefont {Seki}},
  \bibinfo {author} {\bibfnamefont {M.}~\bibnamefont {Kotsugi}},\ and\ \bibinfo
  {author} {\bibfnamefont {K.}~\bibnamefont {Takanashi}},\ }\bibfield  {title}
  {\bibinfo {title} {Magnetic vortex in epitaxially-grown
  {Co}${_2}$({Fe},{Mn}){Si} alloy},\ }\href {https://doi.org/10.1063/1.4945730}
  {\bibfield  {journal} {\bibinfo  {journal} {Applied Physics Letters}\
  }\textbf {\bibinfo {volume} {108}},\ \bibinfo {pages} {152402} (\bibinfo
  {year} {2016}{\natexlab{a}})}\BibitemShut {NoStop}%
\bibitem [{\citenamefont {Yamamoto}\ \emph
  {et~al.}(2016{\natexlab{b}})\citenamefont {Yamamoto}, \citenamefont {Seki},\
  and\ \citenamefont {Takanashi}}]{yamamoto_vortex_2016}%
  \BibitemOpen
  \bibfield  {author} {\bibinfo {author} {\bibfnamefont {T.}~\bibnamefont
  {Yamamoto}}, \bibinfo {author} {\bibfnamefont {T.}~\bibnamefont {Seki}},\
  and\ \bibinfo {author} {\bibfnamefont {K.}~\bibnamefont {Takanashi}},\
  }\bibfield  {title} {\bibinfo {title} {Vortex spin-torque oscillator using
  {Co}${_2}${Fe}${_x}${Mn}${_{1-x}}${Si} {Heusler} alloys},\ }\href
  {https://doi.org/10.1103/PhysRevB.94.094419} {\bibfield  {journal} {\bibinfo
  {journal} {Physical Review B}\ }\textbf {\bibinfo {volume} {94}},\ \bibinfo
  {pages} {094419} (\bibinfo {year} {2016}{\natexlab{b}})}\BibitemShut
  {NoStop}%
\bibitem [{\citenamefont {Seki}\ \emph {et~al.}(2018)\citenamefont {Seki},
  \citenamefont {Kubota}, \citenamefont {Yamamoto},\ and\ \citenamefont
  {Takanashi}}]{seki_size_2018}%
  \BibitemOpen
  \bibfield  {author} {\bibinfo {author} {\bibfnamefont {T.}~\bibnamefont
  {Seki}}, \bibinfo {author} {\bibfnamefont {T.}~\bibnamefont {Kubota}},
  \bibinfo {author} {\bibfnamefont {T.}~\bibnamefont {Yamamoto}},\ and\
  \bibinfo {author} {\bibfnamefont {K.}~\bibnamefont {Takanashi}},\ }\bibfield
  {title} {\bibinfo {title} {Size dependence of vortex-type spin torque
  oscillation in a {Co}${_2}${Fe}${_{0.4}}${Mn}${_{0.6}}${Si} {Heusler} alloy
  disk},\ }\href {https://doi.org/10.1088/1361-6463/aaa748} {\bibfield
  {journal} {\bibinfo  {journal} {Journal of Physics D: Applied Physics}\
  }\textbf {\bibinfo {volume} {51}},\ \bibinfo {pages} {075005} (\bibinfo
  {year} {2018})}\BibitemShut {NoStop}%
\bibitem [{\citenamefont {Guillemard}\ \emph
  {et~al.}(2019{\natexlab{a}})\citenamefont {Guillemard}, \citenamefont
  {Petit-Watelot}, \citenamefont {Rojas-S{\'a}nchez}, \citenamefont {Hohlfeld},
  \citenamefont {Ghanbaja}, \citenamefont {Bataille}, \citenamefont
  {Le~F{\`e}vre}, \citenamefont {Bertran},\ and\ \citenamefont
  {Andrieu}}]{guillemard_polycrystalline_2019}%
  \BibitemOpen
  \bibfield  {author} {\bibinfo {author} {\bibfnamefont {C.}~\bibnamefont
  {Guillemard}}, \bibinfo {author} {\bibfnamefont {S.}~\bibnamefont
  {Petit-Watelot}}, \bibinfo {author} {\bibfnamefont {J.-C.}\ \bibnamefont
  {Rojas-S{\'a}nchez}}, \bibinfo {author} {\bibfnamefont {J.}~\bibnamefont
  {Hohlfeld}}, \bibinfo {author} {\bibfnamefont {J.}~\bibnamefont {Ghanbaja}},
  \bibinfo {author} {\bibfnamefont {A.}~\bibnamefont {Bataille}}, \bibinfo
  {author} {\bibfnamefont {P.}~\bibnamefont {Le~F{\`e}vre}}, \bibinfo {author}
  {\bibfnamefont {F.}~\bibnamefont {Bertran}},\ and\ \bibinfo {author}
  {\bibfnamefont {S.}~\bibnamefont {Andrieu}},\ }\bibfield  {title} {\bibinfo
  {title} {Polycrystalline {Co}${_2}${Mn}-based {Heusler} thin films with high
  spin polarization and low magnetic damping},\ }\href
  {https://doi.org/10.1063/1.5121614} {\bibfield  {journal} {\bibinfo
  {journal} {Applied Physics Letters}\ }\textbf {\bibinfo {volume} {115}},\
  \bibinfo {pages} {172401} (\bibinfo {year} {2019}{\natexlab{a}})}\BibitemShut
  {NoStop}%
\bibitem [{\citenamefont {Guillemard}\ \emph
  {et~al.}(2019{\natexlab{b}})\citenamefont {Guillemard}, \citenamefont
  {Petit-Watelot}, \citenamefont {Pasquier}, \citenamefont {Pierre},
  \citenamefont {Ghanbaja}, \citenamefont {Rojas-S{\'a}nchez}, \citenamefont
  {Bataille}, \citenamefont {Rault}, \citenamefont {Le~F{\`e}vre},
  \citenamefont {Bertran},\ and\ \citenamefont
  {Andrieu}}]{guillemard_ultralow_2019}%
  \BibitemOpen
  \bibfield  {author} {\bibinfo {author} {\bibfnamefont {C.}~\bibnamefont
  {Guillemard}}, \bibinfo {author} {\bibfnamefont {S.}~\bibnamefont
  {Petit-Watelot}}, \bibinfo {author} {\bibfnamefont {L.}~\bibnamefont
  {Pasquier}}, \bibinfo {author} {\bibfnamefont {D.}~\bibnamefont {Pierre}},
  \bibinfo {author} {\bibfnamefont {J.}~\bibnamefont {Ghanbaja}}, \bibinfo
  {author} {\bibfnamefont {J.-C.}\ \bibnamefont {Rojas-S{\'a}nchez}}, \bibinfo
  {author} {\bibfnamefont {A.}~\bibnamefont {Bataille}}, \bibinfo {author}
  {\bibfnamefont {J.}~\bibnamefont {Rault}}, \bibinfo {author} {\bibfnamefont
  {P.}~\bibnamefont {Le~F{\`e}vre}}, \bibinfo {author} {\bibfnamefont
  {F.}~\bibnamefont {Bertran}},\ and\ \bibinfo {author} {\bibfnamefont
  {S.}~\bibnamefont {Andrieu}},\ }\bibfield  {title} {\bibinfo {title}
  {Ultralow magnetic damping in {Co}${_2}${Mn}-based {Heusler} compounds:
  {Promising} materials for spintronics},\ }\href
  {https://doi.org/10.1103/PhysRevApplied.11.064009} {\bibfield  {journal}
  {\bibinfo  {journal} {Physical Review Applied}\ }\textbf {\bibinfo {volume}
  {11}},\ \bibinfo {pages} {064009} (\bibinfo {year}
  {2019}{\natexlab{b}})}\BibitemShut {NoStop}%
\bibitem [{\citenamefont {de~Groot}\ \emph {et~al.}(1983)\citenamefont
  {de~Groot}, \citenamefont {Mueller}, \citenamefont {van Engen},\ and\
  \citenamefont {Buschow}}]{de_groot_new_1983}%
  \BibitemOpen
  \bibfield  {author} {\bibinfo {author} {\bibfnamefont {R.~A.}\ \bibnamefont
  {de~Groot}}, \bibinfo {author} {\bibfnamefont {F.~M.}\ \bibnamefont
  {Mueller}}, \bibinfo {author} {\bibfnamefont {P.~G.}\ \bibnamefont {van
  Engen}},\ and\ \bibinfo {author} {\bibfnamefont {K.~H.~J.}\ \bibnamefont
  {Buschow}},\ }\bibfield  {title} {\bibinfo {title} {New class of materials:
  {Half}-metallic ferromagnets},\ }\href
  {https://doi.org/10.1103/PhysRevLett.50.2024} {\bibfield  {journal} {\bibinfo
   {journal} {Physical Review Letters}\ }\textbf {\bibinfo {volume} {50}},\
  \bibinfo {pages} {2024} (\bibinfo {year} {1983})}\BibitemShut {NoStop}%
\bibitem [{\citenamefont {Ishida}\ \emph {et~al.}(1995)\citenamefont {Ishida},
  \citenamefont {Fujii}, \citenamefont {Kashiwagi},\ and\ \citenamefont
  {Asano}}]{ishida_search_1995}%
  \BibitemOpen
  \bibfield  {author} {\bibinfo {author} {\bibfnamefont {S.}~\bibnamefont
  {Ishida}}, \bibinfo {author} {\bibfnamefont {S.}~\bibnamefont {Fujii}},
  \bibinfo {author} {\bibfnamefont {S.}~\bibnamefont {Kashiwagi}},\ and\
  \bibinfo {author} {\bibfnamefont {S.}~\bibnamefont {Asano}},\ }\bibfield
  {title} {\bibinfo {title} {Search for half-metallic compounds in
  {Co}${_2}${Mn}{Z} ({Z}={IIIb}, {IVb}, {Vb} {element})},\ }\href
  {https://doi.org/10.1143/JPSJ.64.2152} {\bibfield  {journal} {\bibinfo
  {journal} {Journal of the Physical Society of Japan}\ }\textbf {\bibinfo
  {volume} {64}},\ \bibinfo {pages} {2152} (\bibinfo {year}
  {1995})}\BibitemShut {NoStop}%
\bibitem [{\citenamefont {Picozzi}\ \emph {et~al.}(2002)\citenamefont
  {Picozzi}, \citenamefont {Continenza},\ and\ \citenamefont
  {Freeman}}]{picozzi_mathrmco_2mathrmmnx_2002}%
  \BibitemOpen
  \bibfield  {author} {\bibinfo {author} {\bibfnamefont {S.}~\bibnamefont
  {Picozzi}}, \bibinfo {author} {\bibfnamefont {A.}~\bibnamefont
  {Continenza}},\ and\ \bibinfo {author} {\bibfnamefont {A.~J.}\ \bibnamefont
  {Freeman}},\ }\bibfield  {title} {\bibinfo {title} {Co${_2}${Mn}{X}
  ({X}={Si}, {Ge}, {Sn}) {Heusler} compounds: {An} ab initio study of their
  structural, electronic, and magnetic properties at zero and elevated
  pressure},\ }\href {https://doi.org/10.1103/PhysRevB.66.094421} {\bibfield
  {journal} {\bibinfo  {journal} {Physical Review B}\ }\textbf {\bibinfo
  {volume} {66}},\ \bibinfo {pages} {094421} (\bibinfo {year}
  {2002})}\BibitemShut {NoStop}%
\bibitem [{\citenamefont {Kandpal}\ \emph {et~al.}(2007)\citenamefont
  {Kandpal}, \citenamefont {Fecher},\ and\ \citenamefont
  {Felser}}]{kandpal_calculated_2007}%
  \BibitemOpen
  \bibfield  {author} {\bibinfo {author} {\bibfnamefont {H.~C.}\ \bibnamefont
  {Kandpal}}, \bibinfo {author} {\bibfnamefont {G.~H.}\ \bibnamefont
  {Fecher}},\ and\ \bibinfo {author} {\bibfnamefont {C.}~\bibnamefont
  {Felser}},\ }\bibfield  {title} {\bibinfo {title} {Calculated electronic and
  magnetic properties of the half-metallic, transition metal based {Heusler}
  compounds},\ }\href {https://doi.org/10.1088/0022-3727/40/6/S01} {\bibfield
  {journal} {\bibinfo  {journal} {Journal of Physics D: Applied Physics}\
  }\textbf {\bibinfo {volume} {40}},\ \bibinfo {pages} {1507} (\bibinfo {year}
  {2007})}\BibitemShut {NoStop}%
\bibitem [{\citenamefont {Liu}\ \emph {et~al.}(2009)\citenamefont {Liu},
  \citenamefont {Mewes}, \citenamefont {Chshiev}, \citenamefont {Mewes},\ and\
  \citenamefont {Butler}}]{liu_origin_2009}%
  \BibitemOpen
  \bibfield  {author} {\bibinfo {author} {\bibfnamefont {C.}~\bibnamefont
  {Liu}}, \bibinfo {author} {\bibfnamefont {C.~K.~A.}\ \bibnamefont {Mewes}},
  \bibinfo {author} {\bibfnamefont {M.}~\bibnamefont {Chshiev}}, \bibinfo
  {author} {\bibfnamefont {T.}~\bibnamefont {Mewes}},\ and\ \bibinfo {author}
  {\bibfnamefont {W.~H.}\ \bibnamefont {Butler}},\ }\bibfield  {title}
  {\bibinfo {title} {Origin of low {Gilbert} damping in half metals},\ }\href
  {https://doi.org/10.1063/1.3157267} {\bibfield  {journal} {\bibinfo
  {journal} {Applied Physics Letters}\ }\textbf {\bibinfo {volume} {95}},\
  \bibinfo {pages} {022509} (\bibinfo {year} {2009})}\BibitemShut {NoStop}%
\bibitem [{\citenamefont {Thiele}(1973)}]{thiele_steady-state_1973}%
  \BibitemOpen
  \bibfield  {author} {\bibinfo {author} {\bibfnamefont {A.~A.}\ \bibnamefont
  {Thiele}},\ }\bibfield  {title} {\bibinfo {title} {Steady-state motion of
  magnetic domains},\ }\href {https://doi.org/10.1103/PhysRevLett.30.230}
  {\bibfield  {journal} {\bibinfo  {journal} {Physical Review Letters}\
  }\textbf {\bibinfo {volume} {30}},\ \bibinfo {pages} {230} (\bibinfo {year}
  {1973})}\BibitemShut {NoStop}%
\bibitem [{\citenamefont {Petit-Watelot}\ \emph
  {et~al.}(2012{\natexlab{a}})\citenamefont {Petit-Watelot}, \citenamefont
  {Kim}, \citenamefont {Ruotolo}, \citenamefont {Otxoa}, \citenamefont
  {Bouzehouane}, \citenamefont {Grollier}, \citenamefont {Vansteenkiste},
  \citenamefont {Van~de Wiele}, \citenamefont {Cros},\ and\ \citenamefont
  {Devolder}}]{petit-watelot_commensurability_2012}%
  \BibitemOpen
  \bibfield  {author} {\bibinfo {author} {\bibfnamefont {S.}~\bibnamefont
  {Petit-Watelot}}, \bibinfo {author} {\bibfnamefont {J.-V.}\ \bibnamefont
  {Kim}}, \bibinfo {author} {\bibfnamefont {A.}~\bibnamefont {Ruotolo}},
  \bibinfo {author} {\bibfnamefont {R.~M.}\ \bibnamefont {Otxoa}}, \bibinfo
  {author} {\bibfnamefont {K.}~\bibnamefont {Bouzehouane}}, \bibinfo {author}
  {\bibfnamefont {J.}~\bibnamefont {Grollier}}, \bibinfo {author}
  {\bibfnamefont {A.}~\bibnamefont {Vansteenkiste}}, \bibinfo {author}
  {\bibfnamefont {B.}~\bibnamefont {Van~de Wiele}}, \bibinfo {author}
  {\bibfnamefont {V.}~\bibnamefont {Cros}},\ and\ \bibinfo {author}
  {\bibfnamefont {T.}~\bibnamefont {Devolder}},\ }\bibfield  {title} {\bibinfo
  {title} {Commensurability and chaos in magnetic vortex oscillations},\ }\href
  {https://doi.org/10.1038/nphys2362} {\bibfield  {journal} {\bibinfo
  {journal} {Nature Physics}\ }\textbf {\bibinfo {volume} {8}},\ \bibinfo
  {pages} {682} (\bibinfo {year} {2012}{\natexlab{a}})}\BibitemShut {NoStop}%
\bibitem [{\citenamefont {Devolder}\ \emph {et~al.}(2019)\citenamefont
  {Devolder}, \citenamefont {Rontani}, \citenamefont {Petit-Watelot},
  \citenamefont {Bouzehouane}, \citenamefont {Andrieu}, \citenamefont
  {L{\'e}tang}, \citenamefont {Yoo}, \citenamefont {Adam}, \citenamefont
  {Chappert}, \citenamefont {Girod}, \citenamefont {Cros}, \citenamefont
  {Sciamanna},\ and\ \citenamefont {Kim}}]{devolder_chaos_2019}%
  \BibitemOpen
  \bibfield  {author} {\bibinfo {author} {\bibfnamefont {T.}~\bibnamefont
  {Devolder}}, \bibinfo {author} {\bibfnamefont {D.}~\bibnamefont {Rontani}},
  \bibinfo {author} {\bibfnamefont {S.}~\bibnamefont {Petit-Watelot}}, \bibinfo
  {author} {\bibfnamefont {K.}~\bibnamefont {Bouzehouane}}, \bibinfo {author}
  {\bibfnamefont {S.}~\bibnamefont {Andrieu}}, \bibinfo {author} {\bibfnamefont
  {J.}~\bibnamefont {L{\'e}tang}}, \bibinfo {author} {\bibfnamefont {M.-W.}\
  \bibnamefont {Yoo}}, \bibinfo {author} {\bibfnamefont {J.-P.}\ \bibnamefont
  {Adam}}, \bibinfo {author} {\bibfnamefont {C.}~\bibnamefont {Chappert}},
  \bibinfo {author} {\bibfnamefont {S.}~\bibnamefont {Girod}}, \bibinfo
  {author} {\bibfnamefont {V.}~\bibnamefont {Cros}}, \bibinfo {author}
  {\bibfnamefont {M.}~\bibnamefont {Sciamanna}},\ and\ \bibinfo {author}
  {\bibfnamefont {J.-V.}\ \bibnamefont {Kim}},\ }\bibfield  {title} {\bibinfo
  {title} {Chaos in magnetic nanocontact vortex oscillators},\ }\href
  {https://doi.org/10.1103/PhysRevLett.123.147701} {\bibfield  {journal}
  {\bibinfo  {journal} {Physical Review Letters}\ }\textbf {\bibinfo {volume}
  {123}},\ \bibinfo {pages} {147701} (\bibinfo {year} {2019})}\BibitemShut
  {NoStop}%
\bibitem [{\citenamefont {L{\'e}tang}\ \emph {et~al.}(2019)\citenamefont
  {L{\'e}tang}, \citenamefont {Petit-Watelot}, \citenamefont {Yoo},
  \citenamefont {Devolder}, \citenamefont {Bouzehouane}, \citenamefont {Cros},\
  and\ \citenamefont {Kim}}]{letang_modulation_2019}%
  \BibitemOpen
  \bibfield  {author} {\bibinfo {author} {\bibfnamefont {J.}~\bibnamefont
  {L{\'e}tang}}, \bibinfo {author} {\bibfnamefont {S.}~\bibnamefont
  {Petit-Watelot}}, \bibinfo {author} {\bibfnamefont {M.-W.}\ \bibnamefont
  {Yoo}}, \bibinfo {author} {\bibfnamefont {T.}~\bibnamefont {Devolder}},
  \bibinfo {author} {\bibfnamefont {K.}~\bibnamefont {Bouzehouane}}, \bibinfo
  {author} {\bibfnamefont {V.}~\bibnamefont {Cros}},\ and\ \bibinfo {author}
  {\bibfnamefont {J.-V.}\ \bibnamefont {Kim}},\ }\bibfield  {title} {\bibinfo
  {title} {Modulation and phase-locking in nanocontact vortex oscillators},\
  }\href {https://doi.org/10.1103/PhysRevB.100.144414} {\bibfield  {journal}
  {\bibinfo  {journal} {Physical Review B}\ }\textbf {\bibinfo {volume}
  {100}},\ \bibinfo {pages} {144414} (\bibinfo {year} {2019})}\BibitemShut
  {NoStop}%
\bibitem [{\citenamefont {Yoo}\ \emph {et~al.}(2020)\citenamefont {Yoo},
  \citenamefont {Rontani}, \citenamefont {L{\'e}tang}, \citenamefont
  {Petit-Watelot}, \citenamefont {Devolder}, \citenamefont {Sciamanna},
  \citenamefont {Bouzehouane}, \citenamefont {Cros},\ and\ \citenamefont
  {Kim}}]{yoo_pattern_2020}%
  \BibitemOpen
  \bibfield  {author} {\bibinfo {author} {\bibfnamefont {M.-W.}\ \bibnamefont
  {Yoo}}, \bibinfo {author} {\bibfnamefont {D.}~\bibnamefont {Rontani}},
  \bibinfo {author} {\bibfnamefont {J.}~\bibnamefont {L{\'e}tang}}, \bibinfo
  {author} {\bibfnamefont {S.}~\bibnamefont {Petit-Watelot}}, \bibinfo {author}
  {\bibfnamefont {T.}~\bibnamefont {Devolder}}, \bibinfo {author}
  {\bibfnamefont {M.}~\bibnamefont {Sciamanna}}, \bibinfo {author}
  {\bibfnamefont {K.}~\bibnamefont {Bouzehouane}}, \bibinfo {author}
  {\bibfnamefont {V.}~\bibnamefont {Cros}},\ and\ \bibinfo {author}
  {\bibfnamefont {J.-V.}\ \bibnamefont {Kim}},\ }\bibfield  {title} {\bibinfo
  {title} {Pattern generation and symbolic dynamics in a nanocontact vortex
  oscillator},\ }\href {https://doi.org/10.1038/s41467-020-14328-7} {\bibfield
  {journal} {\bibinfo  {journal} {Nature Communications}\ }\textbf {\bibinfo
  {volume} {11}},\ \bibinfo {pages} {601} (\bibinfo {year} {2020})}\BibitemShut
  {NoStop}%
\bibitem [{\citenamefont {Van~Waeyenberge}\ \emph {et~al.}(2006)\citenamefont
  {Van~Waeyenberge}, \citenamefont {Puzic}, \citenamefont {Stoll},
  \citenamefont {Chou}, \citenamefont {Tyliszczak}, \citenamefont {Hertel},
  \citenamefont {F{\"a}hnle}, \citenamefont {Br{\"u}ckl}, \citenamefont {Rott},
  \citenamefont {Reiss}, \citenamefont {Neudecker}, \citenamefont {Weiss},
  \citenamefont {Back},\ and\ \citenamefont
  {Sch{\"u}tz}}]{van_waeyenberge_magnetic_2006}%
  \BibitemOpen
  \bibfield  {author} {\bibinfo {author} {\bibfnamefont {B.}~\bibnamefont
  {Van~Waeyenberge}}, \bibinfo {author} {\bibfnamefont {A.}~\bibnamefont
  {Puzic}}, \bibinfo {author} {\bibfnamefont {H.}~\bibnamefont {Stoll}},
  \bibinfo {author} {\bibfnamefont {K.~W.}\ \bibnamefont {Chou}}, \bibinfo
  {author} {\bibfnamefont {T.}~\bibnamefont {Tyliszczak}}, \bibinfo {author}
  {\bibfnamefont {R.}~\bibnamefont {Hertel}}, \bibinfo {author} {\bibfnamefont
  {M.}~\bibnamefont {F{\"a}hnle}}, \bibinfo {author} {\bibfnamefont
  {H.}~\bibnamefont {Br{\"u}ckl}}, \bibinfo {author} {\bibfnamefont
  {K.}~\bibnamefont {Rott}}, \bibinfo {author} {\bibfnamefont {G.}~\bibnamefont
  {Reiss}}, \bibinfo {author} {\bibfnamefont {I.}~\bibnamefont {Neudecker}},
  \bibinfo {author} {\bibfnamefont {D.}~\bibnamefont {Weiss}}, \bibinfo
  {author} {\bibfnamefont {C.~H.}\ \bibnamefont {Back}},\ and\ \bibinfo
  {author} {\bibfnamefont {G.}~\bibnamefont {Sch{\"u}tz}},\ }\bibfield  {title}
  {\bibinfo {title} {Magnetic vortex core reversal by excitation with short
  bursts of an alternating field},\ }\href
  {https://doi.org/10.1038/nature05240} {\bibfield  {journal} {\bibinfo
  {journal} {Nature}\ }\textbf {\bibinfo {volume} {444}},\ \bibinfo {pages}
  {461} (\bibinfo {year} {2006})}\BibitemShut {NoStop}%
\bibitem [{\citenamefont {Hertel}\ and\ \citenamefont
  {Schneider}(2006)}]{hertel_exchange_2006}%
  \BibitemOpen
  \bibfield  {author} {\bibinfo {author} {\bibfnamefont {R.}~\bibnamefont
  {Hertel}}\ and\ \bibinfo {author} {\bibfnamefont {C.~M.}\ \bibnamefont
  {Schneider}},\ }\bibfield  {title} {\bibinfo {title} {Exchange explosions:
  {Magnetization} dynamics during vortex-antivortex annihilation},\ }\href
  {https://doi.org/10.1103/PhysRevLett.97.177202} {\bibfield  {journal}
  {\bibinfo  {journal} {Physical Review Letters}\ }\textbf {\bibinfo {volume}
  {97}},\ \bibinfo {pages} {177202} (\bibinfo {year} {2006})}\BibitemShut
  {NoStop}%
\bibitem [{\citenamefont {Zhang}\ and\ \citenamefont
  {Li}(2004)}]{zhang_roles_2004}%
  \BibitemOpen
  \bibfield  {author} {\bibinfo {author} {\bibfnamefont {S.}~\bibnamefont
  {Zhang}}\ and\ \bibinfo {author} {\bibfnamefont {Z.}~\bibnamefont {Li}},\
  }\bibfield  {title} {\bibinfo {title} {Roles of nonequilibrium conduction
  electrons on the magnetization dynamics of ferromagnets},\ }\href
  {https://doi.org/10.1103/PhysRevLett.93.127204} {\bibfield  {journal}
  {\bibinfo  {journal} {Physical Review Letters}\ }\textbf {\bibinfo {volume}
  {93}},\ \bibinfo {pages} {127204} (\bibinfo {year} {2004})}\BibitemShut
  {NoStop}%
\bibitem [{\citenamefont
  {Slonczewski}(1996)}]{slonczewski_current-driven_1996}%
  \BibitemOpen
  \bibfield  {author} {\bibinfo {author} {\bibfnamefont {J.}~\bibnamefont
  {Slonczewski}},\ }\bibfield  {title} {\bibinfo {title} {Current-driven
  excitation of magnetic multilayers},\ }\href
  {https://doi.org/10.1016/0304-8853(96)00062-5} {\bibfield  {journal}
  {\bibinfo  {journal} {Journal of Magnetism and Magnetic Materials}\ }\textbf
  {\bibinfo {volume} {159}},\ \bibinfo {pages} {L1} (\bibinfo {year}
  {1996})}\BibitemShut {NoStop}%
\bibitem [{\citenamefont {Pribiag}\ \emph {et~al.}(2007)\citenamefont
  {Pribiag}, \citenamefont {Krivorotov}, \citenamefont {Fuchs}, \citenamefont
  {Braganca}, \citenamefont {Ozatay}, \citenamefont {Sankey}, \citenamefont
  {Ralph},\ and\ \citenamefont {Buhrman}}]{pribiag_magnetic_2007}%
  \BibitemOpen
  \bibfield  {author} {\bibinfo {author} {\bibfnamefont {V.~S.}\ \bibnamefont
  {Pribiag}}, \bibinfo {author} {\bibfnamefont {I.~N.}\ \bibnamefont
  {Krivorotov}}, \bibinfo {author} {\bibfnamefont {G.~D.}\ \bibnamefont
  {Fuchs}}, \bibinfo {author} {\bibfnamefont {P.~M.}\ \bibnamefont {Braganca}},
  \bibinfo {author} {\bibfnamefont {O.}~\bibnamefont {Ozatay}}, \bibinfo
  {author} {\bibfnamefont {J.~C.}\ \bibnamefont {Sankey}}, \bibinfo {author}
  {\bibfnamefont {D.~C.}\ \bibnamefont {Ralph}},\ and\ \bibinfo {author}
  {\bibfnamefont {R.~A.}\ \bibnamefont {Buhrman}},\ }\bibfield  {title}
  {\bibinfo {title} {Magnetic vortex oscillator driven by d.c. spin-polarized
  current},\ }\href {https://doi.org/10.1038/nphys619} {\bibfield  {journal}
  {\bibinfo  {journal} {Nature Physics}\ }\textbf {\bibinfo {volume} {3}},\
  \bibinfo {pages} {498} (\bibinfo {year} {2007})}\BibitemShut {NoStop}%
\bibitem [{\citenamefont {Dussaux}\ \emph {et~al.}(2010)\citenamefont
  {Dussaux}, \citenamefont {Georges}, \citenamefont {Grollier}, \citenamefont
  {Cros}, \citenamefont {Khvalkovskiy}, \citenamefont {Fukushima},
  \citenamefont {Konoto}, \citenamefont {Kubota}, \citenamefont {Yakushiji},
  \citenamefont {Yuasa}, \citenamefont {Zvezdin}, \citenamefont {Ando},\ and\
  \citenamefont {Fert}}]{dussaux_large_2010}%
  \BibitemOpen
  \bibfield  {author} {\bibinfo {author} {\bibfnamefont {A.}~\bibnamefont
  {Dussaux}}, \bibinfo {author} {\bibfnamefont {B.}~\bibnamefont {Georges}},
  \bibinfo {author} {\bibfnamefont {J.}~\bibnamefont {Grollier}}, \bibinfo
  {author} {\bibfnamefont {V.}~\bibnamefont {Cros}}, \bibinfo {author}
  {\bibfnamefont {A.~V.}\ \bibnamefont {Khvalkovskiy}}, \bibinfo {author}
  {\bibfnamefont {A.}~\bibnamefont {Fukushima}}, \bibinfo {author}
  {\bibfnamefont {M.}~\bibnamefont {Konoto}}, \bibinfo {author} {\bibfnamefont
  {H.}~\bibnamefont {Kubota}}, \bibinfo {author} {\bibfnamefont
  {K.}~\bibnamefont {Yakushiji}}, \bibinfo {author} {\bibfnamefont
  {S.}~\bibnamefont {Yuasa}}, \bibinfo {author} {\bibfnamefont {K.~A.}\
  \bibnamefont {Zvezdin}}, \bibinfo {author} {\bibfnamefont {K.}~\bibnamefont
  {Ando}},\ and\ \bibinfo {author} {\bibfnamefont {A.}~\bibnamefont {Fert}},\
  }\bibfield  {title} {\bibinfo {title} {Large microwave generation from
  current-driven magnetic vortex oscillators in magnetic tunnel junctions},\
  }\href {https://doi.org/10.1038/ncomms1006} {\bibfield  {journal} {\bibinfo
  {journal} {Nature Communications}\ }\textbf {\bibinfo {volume} {1}},\
  \bibinfo {pages} {8} (\bibinfo {year} {2010})}\BibitemShut {NoStop}%
\bibitem [{\citenamefont {Locatelli}\ \emph {et~al.}(2011)\citenamefont
  {Locatelli}, \citenamefont {Naletov}, \citenamefont {Grollier}, \citenamefont
  {De~Loubens}, \citenamefont {Cros}, \citenamefont {Deranlot}, \citenamefont
  {Ulysse}, \citenamefont {Faini}, \citenamefont {Klein},\ and\ \citenamefont
  {Fert}}]{Locatelli:2011hw}%
  \BibitemOpen
  \bibfield  {author} {\bibinfo {author} {\bibfnamefont {N.}~\bibnamefont
  {Locatelli}}, \bibinfo {author} {\bibfnamefont {V.~V.}\ \bibnamefont
  {Naletov}}, \bibinfo {author} {\bibfnamefont {J.}~\bibnamefont {Grollier}},
  \bibinfo {author} {\bibfnamefont {G.}~\bibnamefont {De~Loubens}}, \bibinfo
  {author} {\bibfnamefont {V.}~\bibnamefont {Cros}}, \bibinfo {author}
  {\bibfnamefont {C.}~\bibnamefont {Deranlot}}, \bibinfo {author}
  {\bibfnamefont {C.}~\bibnamefont {Ulysse}}, \bibinfo {author} {\bibfnamefont
  {G.}~\bibnamefont {Faini}}, \bibinfo {author} {\bibfnamefont
  {O.}~\bibnamefont {Klein}},\ and\ \bibinfo {author} {\bibfnamefont
  {A.}~\bibnamefont {Fert}},\ }\bibfield  {title} {\bibinfo {title} {{Dynamics
  of two coupled vortices in a spin valve nanopillar excited by spin transfer
  torque}},\ }\href {https://doi.org/10.1063/1.3553771} {\bibfield  {journal}
  {\bibinfo  {journal} {Applied Physics Letters}\ }\textbf {\bibinfo {volume}
  {98}},\ \bibinfo {pages} {062501} (\bibinfo {year} {2011})}\BibitemShut
  {NoStop}%
\bibitem [{\citenamefont {Guillemard}\ \emph {et~al.}(2020)\citenamefont
  {Guillemard}, \citenamefont {Petit-Watelot}, \citenamefont {Devolder},
  \citenamefont {Pasquier}, \citenamefont {Boulet}, \citenamefont {Migot},
  \citenamefont {Ghanbaja}, \citenamefont {Bertran},\ and\ \citenamefont
  {Andrieu}}]{guillemard_issues_2020}%
  \BibitemOpen
  \bibfield  {author} {\bibinfo {author} {\bibfnamefont {C.}~\bibnamefont
  {Guillemard}}, \bibinfo {author} {\bibfnamefont {S.}~\bibnamefont
  {Petit-Watelot}}, \bibinfo {author} {\bibfnamefont {T.}~\bibnamefont
  {Devolder}}, \bibinfo {author} {\bibfnamefont {L.}~\bibnamefont {Pasquier}},
  \bibinfo {author} {\bibfnamefont {P.}~\bibnamefont {Boulet}}, \bibinfo
  {author} {\bibfnamefont {S.}~\bibnamefont {Migot}}, \bibinfo {author}
  {\bibfnamefont {J.}~\bibnamefont {Ghanbaja}}, \bibinfo {author}
  {\bibfnamefont {F.}~\bibnamefont {Bertran}},\ and\ \bibinfo {author}
  {\bibfnamefont {S.}~\bibnamefont {Andrieu}},\ }\bibfield  {title} {\bibinfo
  {title} {Issues in growing {Heusler} compounds in thin films for spintronic
  applications},\ }\href {https://doi.org/10.1063/5.0014241} {\bibfield
  {journal} {\bibinfo  {journal} {Journal of Applied Physics}\ }\textbf
  {\bibinfo {volume} {128}},\ \bibinfo {pages} {241102} (\bibinfo {year}
  {2020})}\BibitemShut {NoStop}%
\bibitem [{\citenamefont {Andrieu}\ \emph {et~al.}(2014)\citenamefont
  {Andrieu}, \citenamefont {Calmels}, \citenamefont {Hauet}, \citenamefont
  {Bonell}, \citenamefont {Le~F{\`e}vre},\ and\ \citenamefont
  {Bertran}}]{andrieu_spectroscopic_2014}%
  \BibitemOpen
  \bibfield  {author} {\bibinfo {author} {\bibfnamefont {S.}~\bibnamefont
  {Andrieu}}, \bibinfo {author} {\bibfnamefont {L.}~\bibnamefont {Calmels}},
  \bibinfo {author} {\bibfnamefont {T.}~\bibnamefont {Hauet}}, \bibinfo
  {author} {\bibfnamefont {F.}~\bibnamefont {Bonell}}, \bibinfo {author}
  {\bibfnamefont {P.}~\bibnamefont {Le~F{\`e}vre}},\ and\ \bibinfo {author}
  {\bibfnamefont {F.}~\bibnamefont {Bertran}},\ }\bibfield  {title} {\bibinfo
  {title} {Spectroscopic and transport studies of
  {Co}${_x}${Fe}${_{1-x}}$/{Mg}{O}(001)-based magnetic tunnel junctions},\
  }\href {https://doi.org/10.1103/PhysRevB.90.214406} {\bibfield  {journal}
  {\bibinfo  {journal} {Physical Review B}\ }\textbf {\bibinfo {volume} {90}},\
  \bibinfo {pages} {214406} (\bibinfo {year} {2014})}\BibitemShut {NoStop}%
\bibitem [{\citenamefont {Nakamura}\ and\ \citenamefont
  {Futamoto}(1993)}]{nakamura_epitaxial_1993}%
  \BibitemOpen
  \bibfield  {author} {\bibinfo {author} {\bibfnamefont {A.}~\bibnamefont
  {Nakamura}}\ and\ \bibinfo {author} {\bibfnamefont {M.}~\bibnamefont
  {Futamoto}},\ }\bibfield  {title} {\bibinfo {title} {Epitaxial {Growth} of
  {Co}/{Cr} {Bilayer} {Films} on {MgO} {Single} {Crystal} {Substrates}},\
  }\href {https://doi.org/10.1143/JJAP.32.L1410} {\bibfield  {journal}
  {\bibinfo  {journal} {Japanese Journal of Applied Physics}\ }\textbf
  {\bibinfo {volume} {32}},\ \bibinfo {pages} {L1410} (\bibinfo {year}
  {1993})}\BibitemShut {NoStop}%
\bibitem [{\citenamefont {Wang}\ \emph {et~al.}(2007)\citenamefont {Wang},
  \citenamefont {Wang}, \citenamefont {Kohn}, \citenamefont {Lee},
  \citenamefont {Goff}, \citenamefont {Singh}, \citenamefont {Barber},\ and\
  \citenamefont {Ward}}]{wang_structural_2007}%
  \BibitemOpen
  \bibfield  {author} {\bibinfo {author} {\bibfnamefont {S.~G.}\ \bibnamefont
  {Wang}}, \bibinfo {author} {\bibfnamefont {C.}~\bibnamefont {Wang}}, \bibinfo
  {author} {\bibfnamefont {A.}~\bibnamefont {Kohn}}, \bibinfo {author}
  {\bibfnamefont {S.}~\bibnamefont {Lee}}, \bibinfo {author} {\bibfnamefont
  {J.~P.}\ \bibnamefont {Goff}}, \bibinfo {author} {\bibfnamefont {L.~J.}\
  \bibnamefont {Singh}}, \bibinfo {author} {\bibfnamefont {Z.~H.}\ \bibnamefont
  {Barber}},\ and\ \bibinfo {author} {\bibfnamefont {R.~C.~C.}\ \bibnamefont
  {Ward}},\ }\bibfield  {title} {\bibinfo {title} {Structural and magnetic
  studies of {Co} layer in epitaxially grown {Fe}/{Co} bilayers},\ }\href
  {https://doi.org/10.1063/1.2672175} {\bibfield  {journal} {\bibinfo
  {journal} {Journal of Applied Physics}\ }\textbf {\bibinfo {volume} {101}},\
  \bibinfo {pages} {09D103} (\bibinfo {year} {2007})}\BibitemShut {NoStop}%
\bibitem [{\citenamefont {Popova}\ \emph {et~al.}(2002)\citenamefont {Popova},
  \citenamefont {Faure-Vincent}, \citenamefont {Tiusan}, \citenamefont
  {Bellouard}, \citenamefont {Fischer}, \citenamefont {Hehn}, \citenamefont
  {Montaigne}, \citenamefont {Alnot}, \citenamefont {Andrieu}, \citenamefont
  {Schuhl}, \citenamefont {Snoeck},\ and\ \citenamefont
  {da~Costa}}]{popova_epitaxial_2002}%
  \BibitemOpen
  \bibfield  {author} {\bibinfo {author} {\bibfnamefont {E.}~\bibnamefont
  {Popova}}, \bibinfo {author} {\bibfnamefont {J.}~\bibnamefont
  {Faure-Vincent}}, \bibinfo {author} {\bibfnamefont {C.}~\bibnamefont
  {Tiusan}}, \bibinfo {author} {\bibfnamefont {C.}~\bibnamefont {Bellouard}},
  \bibinfo {author} {\bibfnamefont {H.}~\bibnamefont {Fischer}}, \bibinfo
  {author} {\bibfnamefont {M.}~\bibnamefont {Hehn}}, \bibinfo {author}
  {\bibfnamefont {F.}~\bibnamefont {Montaigne}}, \bibinfo {author}
  {\bibfnamefont {M.}~\bibnamefont {Alnot}}, \bibinfo {author} {\bibfnamefont
  {S.}~\bibnamefont {Andrieu}}, \bibinfo {author} {\bibfnamefont
  {A.}~\bibnamefont {Schuhl}}, \bibinfo {author} {\bibfnamefont
  {E.}~\bibnamefont {Snoeck}},\ and\ \bibinfo {author} {\bibfnamefont
  {V.}~\bibnamefont {da~Costa}},\ }\bibfield  {title} {\bibinfo {title}
  {Epitaxial {MgO} layer for low-resistance and coupling-free magnetic tunnel
  junctions},\ }\href {https://doi.org/10.1063/1.1498153} {\bibfield  {journal}
  {\bibinfo  {journal} {Applied Physics Letters}\ }\textbf {\bibinfo {volume}
  {81}},\ \bibinfo {pages} {1035} (\bibinfo {year} {2002})}\BibitemShut
  {NoStop}%
\bibitem [{\citenamefont {Gu}\ \emph {et~al.}(1995)\citenamefont {Gu},
  \citenamefont {Gester}, \citenamefont {Hicken}, \citenamefont {Daboo},
  \citenamefont {Tselepi}, \citenamefont {Gray}, \citenamefont {Bland},
  \citenamefont {Brown}, \citenamefont {Thomson},\ and\ \citenamefont
  {Riedi}}]{gu_fourfold_1995}%
  \BibitemOpen
  \bibfield  {author} {\bibinfo {author} {\bibfnamefont {E.}~\bibnamefont
  {Gu}}, \bibinfo {author} {\bibfnamefont {M.}~\bibnamefont {Gester}}, \bibinfo
  {author} {\bibfnamefont {R.~J.}\ \bibnamefont {Hicken}}, \bibinfo {author}
  {\bibfnamefont {C.}~\bibnamefont {Daboo}}, \bibinfo {author} {\bibfnamefont
  {M.}~\bibnamefont {Tselepi}}, \bibinfo {author} {\bibfnamefont {S.~J.}\
  \bibnamefont {Gray}}, \bibinfo {author} {\bibfnamefont {J.~A.~C.}\
  \bibnamefont {Bland}}, \bibinfo {author} {\bibfnamefont {L.~M.}\ \bibnamefont
  {Brown}}, \bibinfo {author} {\bibfnamefont {T.}~\bibnamefont {Thomson}},\
  and\ \bibinfo {author} {\bibfnamefont {P.~C.}\ \bibnamefont {Riedi}},\
  }\bibfield  {title} {\bibinfo {title} {Fourfold anisotropy and structural
  behavior of epitaxial hcp {Co}/{GaAs}(001) thin films},\ }\href
  {https://doi.org/10.1103/PhysRevB.52.14704} {\bibfield  {journal} {\bibinfo
  {journal} {Physical Review B}\ }\textbf {\bibinfo {volume} {52}},\ \bibinfo
  {pages} {14704} (\bibinfo {year} {1995})}\BibitemShut {NoStop}%
\bibitem [{\citenamefont {Bouzehouane}\ \emph {et~al.}(2003)\citenamefont
  {Bouzehouane}, \citenamefont {Fusil}, \citenamefont {Bibes}, \citenamefont
  {Carrey}, \citenamefont {Blon}, \citenamefont {Le~D{\^u}}, \citenamefont
  {Seneor}, \citenamefont {Cros},\ and\ \citenamefont
  {Vila}}]{bouzehouane_nanolithography_2003}%
  \BibitemOpen
  \bibfield  {author} {\bibinfo {author} {\bibfnamefont {K.}~\bibnamefont
  {Bouzehouane}}, \bibinfo {author} {\bibfnamefont {S.}~\bibnamefont {Fusil}},
  \bibinfo {author} {\bibfnamefont {M.}~\bibnamefont {Bibes}}, \bibinfo
  {author} {\bibfnamefont {J.}~\bibnamefont {Carrey}}, \bibinfo {author}
  {\bibfnamefont {T.}~\bibnamefont {Blon}}, \bibinfo {author} {\bibfnamefont
  {M.}~\bibnamefont {Le~D{\^u}}}, \bibinfo {author} {\bibfnamefont
  {P.}~\bibnamefont {Seneor}}, \bibinfo {author} {\bibfnamefont
  {V.}~\bibnamefont {Cros}},\ and\ \bibinfo {author} {\bibfnamefont
  {L.}~\bibnamefont {Vila}},\ }\bibfield  {title} {\bibinfo {title}
  {Nanolithography {based} on {real}-{time} {electrically} {controlled}
  {indentation} with an {atomic} {force} {microscope} for {nanocontact}
  {elaboration}},\ }\href {https://doi.org/10.1021/nl034610j} {\bibfield
  {journal} {\bibinfo  {journal} {Nano Letters}\ }\textbf {\bibinfo {volume}
  {3}},\ \bibinfo {pages} {1599} (\bibinfo {year} {2003})}\BibitemShut
  {NoStop}%
\bibitem [{\citenamefont {Mistral}\ \emph {et~al.}(2008)\citenamefont
  {Mistral}, \citenamefont {van Kampen}, \citenamefont {Hrkac}, \citenamefont
  {Kim}, \citenamefont {Devolder}, \citenamefont {Crozat}, \citenamefont
  {Chappert}, \citenamefont {Lagae},\ and\ \citenamefont
  {Schrefl}}]{mistral_current-driven_2008}%
  \BibitemOpen
  \bibfield  {author} {\bibinfo {author} {\bibfnamefont {Q.}~\bibnamefont
  {Mistral}}, \bibinfo {author} {\bibfnamefont {M.}~\bibnamefont {van Kampen}},
  \bibinfo {author} {\bibfnamefont {G.}~\bibnamefont {Hrkac}}, \bibinfo
  {author} {\bibfnamefont {J.-V.}\ \bibnamefont {Kim}}, \bibinfo {author}
  {\bibfnamefont {T.}~\bibnamefont {Devolder}}, \bibinfo {author}
  {\bibfnamefont {P.}~\bibnamefont {Crozat}}, \bibinfo {author} {\bibfnamefont
  {C.}~\bibnamefont {Chappert}}, \bibinfo {author} {\bibfnamefont
  {L.}~\bibnamefont {Lagae}},\ and\ \bibinfo {author} {\bibfnamefont
  {T.}~\bibnamefont {Schrefl}},\ }\bibfield  {title} {\bibinfo {title}
  {Current-{driven} {vortex} {oscillations} in {metallic} {nanocontacts}},\
  }\href {https://doi.org/10.1103/PhysRevLett.100.257201} {\bibfield  {journal}
  {\bibinfo  {journal} {Physical Review Letters}\ }\textbf {\bibinfo {volume}
  {100}},\ \bibinfo {pages} {257201} (\bibinfo {year} {2008})}\BibitemShut
  {NoStop}%
\bibitem [{\citenamefont {Pufall}\ \emph {et~al.}(2007)\citenamefont {Pufall},
  \citenamefont {Rippard}, \citenamefont {Schneider},\ and\ \citenamefont
  {Russek}}]{pufall_low-field_2007}%
  \BibitemOpen
  \bibfield  {author} {\bibinfo {author} {\bibfnamefont {M.~R.}\ \bibnamefont
  {Pufall}}, \bibinfo {author} {\bibfnamefont {W.~H.}\ \bibnamefont {Rippard}},
  \bibinfo {author} {\bibfnamefont {M.~L.}\ \bibnamefont {Schneider}},\ and\
  \bibinfo {author} {\bibfnamefont {S.~E.}\ \bibnamefont {Russek}},\ }\bibfield
   {title} {\bibinfo {title} {Low-field current-hysteretic oscillations in
  spin-transfer nanocontacts},\ }\href
  {https://doi.org/10.1103/PhysRevB.75.140404} {\bibfield  {journal} {\bibinfo
  {journal} {Physical Review B}\ }\textbf {\bibinfo {volume} {75}},\ \bibinfo
  {pages} {140404(R)} (\bibinfo {year} {2007})}\BibitemShut {NoStop}%
\bibitem [{\citenamefont {Keatley}\ \emph {et~al.}(2016)\citenamefont
  {Keatley}, \citenamefont {Sani}, \citenamefont {Hrkac}, \citenamefont
  {Mohseni}, \citenamefont {D{\"u}rrenfeld}, \citenamefont {Loughran},
  \citenamefont {{\AA}kerman},\ and\ \citenamefont
  {Hicken}}]{keatley_direct_2016}%
  \BibitemOpen
  \bibfield  {author} {\bibinfo {author} {\bibfnamefont {P.~S.}\ \bibnamefont
  {Keatley}}, \bibinfo {author} {\bibfnamefont {S.~R.}\ \bibnamefont {Sani}},
  \bibinfo {author} {\bibfnamefont {G.}~\bibnamefont {Hrkac}}, \bibinfo
  {author} {\bibfnamefont {S.~M.}\ \bibnamefont {Mohseni}}, \bibinfo {author}
  {\bibfnamefont {P.}~\bibnamefont {D{\"u}rrenfeld}}, \bibinfo {author}
  {\bibfnamefont {T.~H.~J.}\ \bibnamefont {Loughran}}, \bibinfo {author}
  {\bibfnamefont {J.}~\bibnamefont {{\AA}kerman}},\ and\ \bibinfo {author}
  {\bibfnamefont {R.~J.}\ \bibnamefont {Hicken}},\ }\bibfield  {title}
  {\bibinfo {title} {Direct observation of magnetization dynamics generated by
  nanocontact spin-torque vortex oscillators},\ }\href
  {https://doi.org/10.1103/PhysRevB.94.060402} {\bibfield  {journal} {\bibinfo
  {journal} {Physical Review B}\ }\textbf {\bibinfo {volume} {94}},\ \bibinfo
  {pages} {060402(R)} (\bibinfo {year} {2016})}\BibitemShut {NoStop}%
\bibitem [{\citenamefont {Krivorotov}\ \emph {et~al.}(2007)\citenamefont
  {Krivorotov}, \citenamefont {Berkov}, \citenamefont {Gorn}, \citenamefont
  {Emley}, \citenamefont {Sankey}, \citenamefont {Ralph},\ and\ \citenamefont
  {Buhrman}}]{krivorotov_large-amplitude_2007}%
  \BibitemOpen
  \bibfield  {author} {\bibinfo {author} {\bibfnamefont {I.~N.}\ \bibnamefont
  {Krivorotov}}, \bibinfo {author} {\bibfnamefont {D.~V.}\ \bibnamefont
  {Berkov}}, \bibinfo {author} {\bibfnamefont {N.~L.}\ \bibnamefont {Gorn}},
  \bibinfo {author} {\bibfnamefont {N.~C.}\ \bibnamefont {Emley}}, \bibinfo
  {author} {\bibfnamefont {J.~C.}\ \bibnamefont {Sankey}}, \bibinfo {author}
  {\bibfnamefont {D.~C.}\ \bibnamefont {Ralph}},\ and\ \bibinfo {author}
  {\bibfnamefont {R.~A.}\ \bibnamefont {Buhrman}},\ }\bibfield  {title}
  {\bibinfo {title} {Large-amplitude coherent spin waves excited by
  spin-polarized current in nanoscale spin valves},\ }\href
  {https://doi.org/10.1103/PhysRevB.76.024418} {\bibfield  {journal} {\bibinfo
  {journal} {Physical Review B}\ }\textbf {\bibinfo {volume} {76}},\ \bibinfo
  {pages} {024418} (\bibinfo {year} {2007})}\BibitemShut {NoStop}%
\bibitem [{\citenamefont {Pribiag}\ \emph {et~al.}(2009)\citenamefont
  {Pribiag}, \citenamefont {Finocchio}, \citenamefont {Williams}, \citenamefont
  {Ralph},\ and\ \citenamefont {Buhrman}}]{pribiag_long-timescale_2009}%
  \BibitemOpen
  \bibfield  {author} {\bibinfo {author} {\bibfnamefont {V.~S.}\ \bibnamefont
  {Pribiag}}, \bibinfo {author} {\bibfnamefont {G.}~\bibnamefont {Finocchio}},
  \bibinfo {author} {\bibfnamefont {B.~J.}\ \bibnamefont {Williams}}, \bibinfo
  {author} {\bibfnamefont {D.~C.}\ \bibnamefont {Ralph}},\ and\ \bibinfo
  {author} {\bibfnamefont {R.~A.}\ \bibnamefont {Buhrman}},\ }\bibfield
  {title} {\bibinfo {title} {Long-timescale fluctuations in zero-field magnetic
  vortex oscillations driven by dc spin-polarized current},\ }\href
  {https://doi.org/10.1103/PhysRevB.80.180411} {\bibfield  {journal} {\bibinfo
  {journal} {Physical Review B}\ }\textbf {\bibinfo {volume} {80}},\ \bibinfo
  {pages} {180411(R)} (\bibinfo {year} {2009})}\BibitemShut {NoStop}%
\bibitem [{\citenamefont {Kuepferling}\ \emph {et~al.}(2010)\citenamefont
  {Kuepferling}, \citenamefont {Serpico}, \citenamefont {Pufall}, \citenamefont
  {Rippard}, \citenamefont {Wallis}, \citenamefont {Imtiaz}, \citenamefont
  {Krivosik}, \citenamefont {Pasquale},\ and\ \citenamefont
  {Kabos}}]{kuepferling_two_2010}%
  \BibitemOpen
  \bibfield  {author} {\bibinfo {author} {\bibfnamefont {M.}~\bibnamefont
  {Kuepferling}}, \bibinfo {author} {\bibfnamefont {C.}~\bibnamefont
  {Serpico}}, \bibinfo {author} {\bibfnamefont {M.}~\bibnamefont {Pufall}},
  \bibinfo {author} {\bibfnamefont {W.}~\bibnamefont {Rippard}}, \bibinfo
  {author} {\bibfnamefont {T.~M.}\ \bibnamefont {Wallis}}, \bibinfo {author}
  {\bibfnamefont {A.}~\bibnamefont {Imtiaz}}, \bibinfo {author} {\bibfnamefont
  {P.}~\bibnamefont {Krivosik}}, \bibinfo {author} {\bibfnamefont
  {M.}~\bibnamefont {Pasquale}},\ and\ \bibinfo {author} {\bibfnamefont
  {P.}~\bibnamefont {Kabos}},\ }\bibfield  {title} {\bibinfo {title} {Two modes
  behavior of vortex oscillations in spin-transfer nanocontacts subject to
  in-plane magnetic fields},\ }\href {https://doi.org/10.1063/1.3455883}
  {\bibfield  {journal} {\bibinfo  {journal} {Applied Physics Letters}\
  }\textbf {\bibinfo {volume} {96}},\ \bibinfo {pages} {252507} (\bibinfo
  {year} {2010})}\BibitemShut {NoStop}%
\bibitem [{\citenamefont {Eggeling}\ \emph {et~al.}(2011)\citenamefont
  {Eggeling}, \citenamefont {Dimopoulos}, \citenamefont {Uhrmann},
  \citenamefont {Bethge}, \citenamefont {Heer}, \citenamefont {H{\"o}ink},\
  and\ \citenamefont {Br{\"u}ckl}}]{eggeling_low_2011}%
  \BibitemOpen
  \bibfield  {author} {\bibinfo {author} {\bibfnamefont {M.}~\bibnamefont
  {Eggeling}}, \bibinfo {author} {\bibfnamefont {T.}~\bibnamefont
  {Dimopoulos}}, \bibinfo {author} {\bibfnamefont {T.}~\bibnamefont {Uhrmann}},
  \bibinfo {author} {\bibfnamefont {O.}~\bibnamefont {Bethge}}, \bibinfo
  {author} {\bibfnamefont {R.}~\bibnamefont {Heer}}, \bibinfo {author}
  {\bibfnamefont {V.}~\bibnamefont {H{\"o}ink}},\ and\ \bibinfo {author}
  {\bibfnamefont {H.}~\bibnamefont {Br{\"u}ckl}},\ }\bibfield  {title}
  {\bibinfo {title} {Low spin current-driven dynamic excitations and
  metastability in spin-valve nanocontacts with unpinned artificial
  antiferromagnet},\ }\href {https://doi.org/10.1063/1.3537953} {\bibfield
  {journal} {\bibinfo  {journal} {Applied Physics Letters}\ }\textbf {\bibinfo
  {volume} {98}},\ \bibinfo {pages} {042504} (\bibinfo {year}
  {2011})}\BibitemShut {NoStop}%
\bibitem [{\citenamefont {Wang}\ \emph
  {et~al.}(2011{\natexlab{a}})\citenamefont {Wang}, \citenamefont {Wang},
  \citenamefont {Qin}, \citenamefont {Yeung}, \citenamefont {Kwok},
  \citenamefont {Wong}, \citenamefont {Xue}, \citenamefont {Chu}, \citenamefont
  {Leung},\ and\ \citenamefont {Ruotolo}}]{wang_multiple-mode_2011}%
  \BibitemOpen
  \bibfield  {author} {\bibinfo {author} {\bibfnamefont {N.}~\bibnamefont
  {Wang}}, \bibinfo {author} {\bibfnamefont {X.~L.}\ \bibnamefont {Wang}},
  \bibinfo {author} {\bibfnamefont {W.}~\bibnamefont {Qin}}, \bibinfo {author}
  {\bibfnamefont {S.~H.}\ \bibnamefont {Yeung}}, \bibinfo {author}
  {\bibfnamefont {D.~T.~K.}\ \bibnamefont {Kwok}}, \bibinfo {author}
  {\bibfnamefont {H.~F.}\ \bibnamefont {Wong}}, \bibinfo {author}
  {\bibfnamefont {Q.}~\bibnamefont {Xue}}, \bibinfo {author} {\bibfnamefont
  {P.~K.}\ \bibnamefont {Chu}}, \bibinfo {author} {\bibfnamefont {C.~W.}\
  \bibnamefont {Leung}},\ and\ \bibinfo {author} {\bibfnamefont
  {A.}~\bibnamefont {Ruotolo}},\ }\bibfield  {title} {\bibinfo {title}
  {Multiple-mode excitation in spin-transfer nanocontacts with dynamic
  polarizer},\ }\href {https://doi.org/10.1063/1.3600328} {\bibfield  {journal}
  {\bibinfo  {journal} {Applied Physics Letters}\ }\textbf {\bibinfo {volume}
  {98}},\ \bibinfo {pages} {242506} (\bibinfo {year}
  {2011}{\natexlab{a}})}\BibitemShut {NoStop}%
\bibitem [{\citenamefont {Keatley}\ \emph {et~al.}(2017)\citenamefont
  {Keatley}, \citenamefont {Sani}, \citenamefont {Hrkac}, \citenamefont
  {Mohseni}, \citenamefont {D{\"u}rrenfeld}, \citenamefont {{\AA}kerman},\ and\
  \citenamefont {Hicken}}]{keatley_imaging_2017}%
  \BibitemOpen
  \bibfield  {author} {\bibinfo {author} {\bibfnamefont {P.~S.}\ \bibnamefont
  {Keatley}}, \bibinfo {author} {\bibfnamefont {S.~R.}\ \bibnamefont {Sani}},
  \bibinfo {author} {\bibfnamefont {G.}~\bibnamefont {Hrkac}}, \bibinfo
  {author} {\bibfnamefont {S.~M.}\ \bibnamefont {Mohseni}}, \bibinfo {author}
  {\bibfnamefont {P.}~\bibnamefont {D{\"u}rrenfeld}}, \bibinfo {author}
  {\bibfnamefont {J.}~\bibnamefont {{\AA}kerman}},\ and\ \bibinfo {author}
  {\bibfnamefont {R.~J.}\ \bibnamefont {Hicken}},\ }\bibfield  {title}
  {\bibinfo {title} {Imaging magnetisation dynamics in nano-contact spin-torque
  vortex oscillators exhibiting gyrotropic mode splitting},\ }\href
  {https://doi.org/10.1088/1361-6463/aa628a} {\bibfield  {journal} {\bibinfo
  {journal} {Journal of Physics D: Applied Physics}\ }\textbf {\bibinfo
  {volume} {50}},\ \bibinfo {pages} {164003} (\bibinfo {year}
  {2017})}\BibitemShut {NoStop}%
\bibitem [{\citenamefont {Zhang}\ \emph {et~al.}(2017)\citenamefont {Zhang},
  \citenamefont {Iacocca},\ and\ \citenamefont
  {Heinonen}}]{zhang_tunable_2017}%
  \BibitemOpen
  \bibfield  {author} {\bibinfo {author} {\bibfnamefont {S.~S.-L.}\
  \bibnamefont {Zhang}}, \bibinfo {author} {\bibfnamefont {E.}~\bibnamefont
  {Iacocca}},\ and\ \bibinfo {author} {\bibfnamefont {O.}~\bibnamefont
  {Heinonen}},\ }\bibfield  {title} {\bibinfo {title} {Tunable {mode}
  {coupling} in {nanocontact} {spin}-{torque} {oscillators}},\ }\href
  {https://doi.org/10.1103/PhysRevApplied.8.014034} {\bibfield  {journal}
  {\bibinfo  {journal} {Physical Review Applied}\ }\textbf {\bibinfo {volume}
  {8}},\ \bibinfo {pages} {014034} (\bibinfo {year} {2017})}\BibitemShut
  {NoStop}%
\bibitem [{\citenamefont {Martin}\ \emph {et~al.}(2013)\citenamefont {Martin},
  \citenamefont {Thirion}, \citenamefont {Hoarau}, \citenamefont {Baraduc},\
  and\ \citenamefont {Di{\'e}ny}}]{martin_tunability_2013}%
  \BibitemOpen
  \bibfield  {author} {\bibinfo {author} {\bibfnamefont {S.~Y.}\ \bibnamefont
  {Martin}}, \bibinfo {author} {\bibfnamefont {C.}~\bibnamefont {Thirion}},
  \bibinfo {author} {\bibfnamefont {C.}~\bibnamefont {Hoarau}}, \bibinfo
  {author} {\bibfnamefont {C.}~\bibnamefont {Baraduc}},\ and\ \bibinfo {author}
  {\bibfnamefont {B.}~\bibnamefont {Di{\'e}ny}},\ }\bibfield  {title} {\bibinfo
  {title} {Tunability versus deviation sensitivity in a nonlinear vortex
  oscillator},\ }\href {https://doi.org/10.1103/PhysRevB.88.024421} {\bibfield
  {journal} {\bibinfo  {journal} {Physical Review B}\ }\textbf {\bibinfo
  {volume} {88}},\ \bibinfo {pages} {024421} (\bibinfo {year}
  {2013})}\BibitemShut {NoStop}%
\bibitem [{\citenamefont {Soucaille}\ \emph {et~al.}(2017)\citenamefont
  {Soucaille}, \citenamefont {Kim}, \citenamefont {Devolder}, \citenamefont
  {Petit-Watelot}, \citenamefont {Manfrini}, \citenamefont {Roy},\ and\
  \citenamefont {Lagae}}]{soucaille_nanocontact_2017}%
  \BibitemOpen
  \bibfield  {author} {\bibinfo {author} {\bibfnamefont {R.}~\bibnamefont
  {Soucaille}}, \bibinfo {author} {\bibfnamefont {J.-V.}\ \bibnamefont {Kim}},
  \bibinfo {author} {\bibfnamefont {T.}~\bibnamefont {Devolder}}, \bibinfo
  {author} {\bibfnamefont {S.}~\bibnamefont {Petit-Watelot}}, \bibinfo {author}
  {\bibfnamefont {M.}~\bibnamefont {Manfrini}}, \bibinfo {author}
  {\bibfnamefont {W.~V.}\ \bibnamefont {Roy}},\ and\ \bibinfo {author}
  {\bibfnamefont {L.}~\bibnamefont {Lagae}},\ }\bibfield  {title} {\bibinfo
  {title} {Nanocontact based spin torque oscillators with two free layers},\
  }\href {https://doi.org/10.1088/1361-6463/aa5627} {\bibfield  {journal}
  {\bibinfo  {journal} {Journal of Physics D: Applied Physics}\ }\textbf
  {\bibinfo {volume} {50}},\ \bibinfo {pages} {085002} (\bibinfo {year}
  {2017})}\BibitemShut {NoStop}%
\bibitem [{\citenamefont {Welch}(1967)}]{welch_use_1967}%
  \BibitemOpen
  \bibfield  {author} {\bibinfo {author} {\bibfnamefont {P.}~\bibnamefont
  {Welch}},\ }\bibfield  {title} {\bibinfo {title} {The use of fast {Fourier}
  transform for the estimation of power spectra: {A} method based on time
  averaging over short, modified periodograms},\ }\href
  {https://doi.org/10.1109/TAU.1967.1161901} {\bibfield  {journal} {\bibinfo
  {journal} {IEEE Transactions on Audio and Electroacoustics}\ }\textbf
  {\bibinfo {volume} {15}},\ \bibinfo {pages} {70} (\bibinfo {year}
  {1967})}\BibitemShut {NoStop}%
\bibitem [{\citenamefont {Poon}\ and\ \citenamefont
  {Barahona}(2001)}]{poon_titration_2001}%
  \BibitemOpen
  \bibfield  {author} {\bibinfo {author} {\bibfnamefont {C.-S.}\ \bibnamefont
  {Poon}}\ and\ \bibinfo {author} {\bibfnamefont {M.}~\bibnamefont
  {Barahona}},\ }\bibfield  {title} {\bibinfo {title} {Titration of chaos with
  added noise},\ }\href {https://doi.org/10.1073/pnas.131173198} {\bibfield
  {journal} {\bibinfo  {journal} {Proceedings of the National Academy of
  Sciences}\ }\textbf {\bibinfo {volume} {98}},\ \bibinfo {pages} {7107}
  (\bibinfo {year} {2001})}\BibitemShut {NoStop}%
\bibitem [{\citenamefont {Hu}\ and\ \citenamefont
  {Raman}(2006)}]{hu_chaos_2006}%
  \BibitemOpen
  \bibfield  {author} {\bibinfo {author} {\bibfnamefont {S.}~\bibnamefont
  {Hu}}\ and\ \bibinfo {author} {\bibfnamefont {A.}~\bibnamefont {Raman}},\
  }\bibfield  {title} {\bibinfo {title} {Chaos in {atomic} {force}
  {microscopy}},\ }\href {https://doi.org/10.1103/PhysRevLett.96.036107}
  {\bibfield  {journal} {\bibinfo  {journal} {Physical Review Letters}\
  }\textbf {\bibinfo {volume} {96}},\ \bibinfo {pages} {036107} (\bibinfo
  {year} {2006})}\BibitemShut {NoStop}%
\bibitem [{\citenamefont {Kantz}\ and\ \citenamefont
  {Schreiber}(2003)}]{kantz_nonlinear_2003}%
  \BibitemOpen
  \bibfield  {author} {\bibinfo {author} {\bibfnamefont {H.}~\bibnamefont
  {Kantz}}\ and\ \bibinfo {author} {\bibfnamefont {T.}~\bibnamefont
  {Schreiber}},\ }\href@noop {} {\emph {\bibinfo {title} {Nonlinear {Time}
  {Series} {Analysis}}}},\ \bibinfo {edition} {2nd}\ ed.\ (\bibinfo
  {publisher} {Cambridge University Press},\ \bibinfo {address} {Cambridge,
  England},\ \bibinfo {year} {2003})\BibitemShut {NoStop}%
\bibitem [{\citenamefont {Barahona}\ and\ \citenamefont
  {Poon}(1997)}]{Poon_NLD_1997}%
  \BibitemOpen
  \bibfield  {author} {\bibinfo {author} {\bibfnamefont {M.}~\bibnamefont
  {Barahona}}\ and\ \bibinfo {author} {\bibfnamefont {C.-S.}\ \bibnamefont
  {Poon}},\ }\bibfield  {title} {\bibinfo {title} {Detection of nonlinear
  dynamics in short, noisy time series},\ }\href
  {https://doi.org/10.1038/381215a0} {\bibfield  {journal} {\bibinfo  {journal}
  {Nature (London)}\ }\textbf {\bibinfo {volume} {381}},\ \bibinfo {pages}
  {215} (\bibinfo {year} {1997})}\BibitemShut {NoStop}%
\bibitem [{\citenamefont {Tretiakov}\ and\ \citenamefont
  {Tchernyshyov}(2007)}]{tretiakov_vortices_2007}%
  \BibitemOpen
  \bibfield  {author} {\bibinfo {author} {\bibfnamefont {O.~A.}\ \bibnamefont
  {Tretiakov}}\ and\ \bibinfo {author} {\bibfnamefont {O.}~\bibnamefont
  {Tchernyshyov}},\ }\bibfield  {title} {\bibinfo {title} {Vortices in thin
  ferromagnetic films and the skyrmion number},\ }\href
  {https://doi.org/10.1103/PhysRevB.75.012408} {\bibfield  {journal} {\bibinfo
  {journal} {Physical Review B}\ }\textbf {\bibinfo {volume} {75}},\ \bibinfo
  {pages} {012408} (\bibinfo {year} {2007})}\BibitemShut {NoStop}%
\bibitem [{\citenamefont {Guslienko}\ \emph {et~al.}(2008)\citenamefont
  {Guslienko}, \citenamefont {Lee},\ and\ \citenamefont
  {Kim}}]{guslienko_dynamic_2008}%
  \BibitemOpen
  \bibfield  {author} {\bibinfo {author} {\bibfnamefont {K.~Y.}\ \bibnamefont
  {Guslienko}}, \bibinfo {author} {\bibfnamefont {K.-S.}\ \bibnamefont {Lee}},\
  and\ \bibinfo {author} {\bibfnamefont {S.-K.}\ \bibnamefont {Kim}},\
  }\bibfield  {title} {\bibinfo {title} {Dynamic {origin} of {vortex} {core}
  {switching} in {soft} {magnetic} {nanodots}},\ }\href
  {https://doi.org/10.1103/PhysRevLett.100.027203} {\bibfield  {journal}
  {\bibinfo  {journal} {Physical Review Letters}\ }\textbf {\bibinfo {volume}
  {100}},\ \bibinfo {pages} {027203} (\bibinfo {year} {2008})}\BibitemShut
  {NoStop}%
\bibitem [{\citenamefont {Yoo}\ \emph {et~al.}(2021)\citenamefont {Yoo},
  \citenamefont {Mineo},\ and\ \citenamefont {Kim}}]{yoo_analytical_2021}%
  \BibitemOpen
  \bibfield  {author} {\bibinfo {author} {\bibfnamefont {M.-W.}\ \bibnamefont
  {Yoo}}, \bibinfo {author} {\bibfnamefont {F.}~\bibnamefont {Mineo}},\ and\
  \bibinfo {author} {\bibfnamefont {J.-V.}\ \bibnamefont {Kim}},\ }\bibfield
  {title} {\bibinfo {title} {Analytical model of the deformation-induced
  inertial dynamics of a magnetic vortex},\ }\href
  {https://doi.org/10.1063/5.0039535} {\bibfield  {journal} {\bibinfo
  {journal} {Journal of Applied Physics}\ }\textbf {\bibinfo {volume} {129}},\
  \bibinfo {pages} {053903} (\bibinfo {year} {2021})}\BibitemShut {NoStop}%
\bibitem [{\citenamefont {Muduli}\ \emph {et~al.}(2012)\citenamefont {Muduli},
  \citenamefont {Heinonen},\ and\ \citenamefont
  {{\AA}kerman}}]{muduli_decoherence_2012}%
  \BibitemOpen
  \bibfield  {author} {\bibinfo {author} {\bibfnamefont {P.~K.}\ \bibnamefont
  {Muduli}}, \bibinfo {author} {\bibfnamefont {O.~G.}\ \bibnamefont
  {Heinonen}},\ and\ \bibinfo {author} {\bibfnamefont {J.}~\bibnamefont
  {{\AA}kerman}},\ }\bibfield  {title} {\bibinfo {title} {Decoherence and
  {mode} {hopping} in a {magnetic} {tunnel} {junction} {based} {spin} {torque}
  {oscillator}},\ }\href {https://doi.org/10.1103/PhysRevLett.108.207203}
  {\bibfield  {journal} {\bibinfo  {journal} {Physical Review Letters}\
  }\textbf {\bibinfo {volume} {108}},\ \bibinfo {pages} {207203} (\bibinfo
  {year} {2012})}\BibitemShut {NoStop}%
\bibitem [{\citenamefont {Heinonen}\ \emph {et~al.}(2013)\citenamefont
  {Heinonen}, \citenamefont {Muduli}, \citenamefont {Iacocca},\ and\
  \citenamefont {{\AA}kerman}}]{Heinonen:2013iu}%
  \BibitemOpen
  \bibfield  {author} {\bibinfo {author} {\bibfnamefont {O.~G.}\ \bibnamefont
  {Heinonen}}, \bibinfo {author} {\bibfnamefont {P.~K.}\ \bibnamefont
  {Muduli}}, \bibinfo {author} {\bibfnamefont {E.}~\bibnamefont {Iacocca}},\
  and\ \bibinfo {author} {\bibfnamefont {J.}~\bibnamefont {{\AA}kerman}},\
  }\bibfield  {title} {\bibinfo {title} {Decoherence, mode hopping, and mode
  coupling in spin torque oscillators},\ }\href
  {https://doi.org/10.1109/TMAG.2013.2242866} {\bibfield  {journal} {\bibinfo
  {journal} {IEEE Transactions on Magnetics}\ }\textbf {\bibinfo {volume}
  {49}},\ \bibinfo {pages} {4398} (\bibinfo {year} {2013})}\BibitemShut
  {NoStop}%
\bibitem [{\citenamefont {Dumas}\ \emph {et~al.}(2013)\citenamefont {Dumas},
  \citenamefont {Iacocca}, \citenamefont {Bonetti}, \citenamefont {Sani},
  \citenamefont {Mohseni}, \citenamefont {Eklund}, \citenamefont {Persson},
  \citenamefont {Heinonen},\ and\ \citenamefont {{\AA}kerman}}]{Dumas:2013bx}%
  \BibitemOpen
  \bibfield  {author} {\bibinfo {author} {\bibfnamefont {R.~K.}\ \bibnamefont
  {Dumas}}, \bibinfo {author} {\bibfnamefont {E.}~\bibnamefont {Iacocca}},
  \bibinfo {author} {\bibfnamefont {S.}~\bibnamefont {Bonetti}}, \bibinfo
  {author} {\bibfnamefont {S.~R.}\ \bibnamefont {Sani}}, \bibinfo {author}
  {\bibfnamefont {S.~M.}\ \bibnamefont {Mohseni}}, \bibinfo {author}
  {\bibfnamefont {A.}~\bibnamefont {Eklund}}, \bibinfo {author} {\bibfnamefont
  {J.}~\bibnamefont {Persson}}, \bibinfo {author} {\bibfnamefont
  {O.}~\bibnamefont {Heinonen}},\ and\ \bibinfo {author} {\bibfnamefont
  {J.}~\bibnamefont {{\AA}kerman}},\ }\bibfield  {title} {\bibinfo {title}
  {Spin-wave-mode coexistence on the nanoscale: A consequence of the
  oersted-field-induced asymmetric energy landscape},\ }\href
  {https://doi.org/10.1103/PhysRevLett.110.257202} {\bibfield  {journal}
  {\bibinfo  {journal} {Physical Review Letters}\ }\textbf {\bibinfo {volume}
  {110}},\ \bibinfo {pages} {257202} (\bibinfo {year} {2013})}\BibitemShut
  {NoStop}%
\bibitem [{\citenamefont {Iacocca}\ \emph {et~al.}(2015)\citenamefont
  {Iacocca}, \citenamefont {D{\"u}rrenfeld}, \citenamefont {Heinonen},
  \citenamefont {{\AA}kerman},\ and\ \citenamefont {Dumas}}]{Iacocca:2015ev}%
  \BibitemOpen
  \bibfield  {author} {\bibinfo {author} {\bibfnamefont {E.}~\bibnamefont
  {Iacocca}}, \bibinfo {author} {\bibfnamefont {P.}~\bibnamefont
  {D{\"u}rrenfeld}}, \bibinfo {author} {\bibfnamefont {O.}~\bibnamefont
  {Heinonen}}, \bibinfo {author} {\bibfnamefont {J.}~\bibnamefont
  {{\AA}kerman}},\ and\ \bibinfo {author} {\bibfnamefont {R.~K.}\ \bibnamefont
  {Dumas}},\ }\bibfield  {title} {\bibinfo {title} {{Mode-coupling mechanisms
  in nanocontact spin-torque oscillators}},\ }\href
  {https://doi.org/10.1103/PhysRevB.91.104405} {\bibfield  {journal} {\bibinfo
  {journal} {Physical Review B}\ }\textbf {\bibinfo {volume} {91}},\ \bibinfo
  {pages} {104405} (\bibinfo {year} {2015})}\BibitemShut {NoStop}%
\bibitem [{\citenamefont {Wang}\ \emph
  {et~al.}(2011{\natexlab{b}})\citenamefont {Wang}, \citenamefont {Wang},
  \citenamefont {Kwok},\ and\ \citenamefont {Ruotolo}}]{wang_three-mode_2011}%
  \BibitemOpen
  \bibfield  {author} {\bibinfo {author} {\bibfnamefont {N.}~\bibnamefont
  {Wang}}, \bibinfo {author} {\bibfnamefont {X.~L.}\ \bibnamefont {Wang}},
  \bibinfo {author} {\bibfnamefont {D.~T.~K.}\ \bibnamefont {Kwok}},\ and\
  \bibinfo {author} {\bibfnamefont {A.}~\bibnamefont {Ruotolo}},\ }\bibfield
  {title} {\bibinfo {title} {Three-{mode} {behavior} of {spin}-{transfer}
  {vortex} {oscillators} {with} {dynamic} {polarizer}},\ }\href
  {https://doi.org/10.1109/TMAG.2011.2158070} {\bibfield  {journal} {\bibinfo
  {journal} {IEEE Transactions on Magnetics}\ }\textbf {\bibinfo {volume}
  {47}},\ \bibinfo {pages} {3704} (\bibinfo {year}
  {2011}{\natexlab{b}})}\BibitemShut {NoStop}%
\bibitem [{\citenamefont {Lebrun}\ \emph {et~al.}(2014)\citenamefont {Lebrun},
  \citenamefont {Locatelli}, \citenamefont {Tsunegi}, \citenamefont {Grollier},
  \citenamefont {Cros}, \citenamefont {Abreu~Araujo}, \citenamefont {Kubota},
  \citenamefont {Yakushiji}, \citenamefont {Fukushima},\ and\ \citenamefont
  {Yuasa}}]{lebrun_nonlinear_2014}%
  \BibitemOpen
  \bibfield  {author} {\bibinfo {author} {\bibfnamefont {R.}~\bibnamefont
  {Lebrun}}, \bibinfo {author} {\bibfnamefont {N.}~\bibnamefont {Locatelli}},
  \bibinfo {author} {\bibfnamefont {S.}~\bibnamefont {Tsunegi}}, \bibinfo
  {author} {\bibfnamefont {J.}~\bibnamefont {Grollier}}, \bibinfo {author}
  {\bibfnamefont {V.}~\bibnamefont {Cros}}, \bibinfo {author} {\bibfnamefont
  {F.}~\bibnamefont {Abreu~Araujo}}, \bibinfo {author} {\bibfnamefont
  {H.}~\bibnamefont {Kubota}}, \bibinfo {author} {\bibfnamefont
  {K.}~\bibnamefont {Yakushiji}}, \bibinfo {author} {\bibfnamefont
  {A.}~\bibnamefont {Fukushima}},\ and\ \bibinfo {author} {\bibfnamefont
  {S.}~\bibnamefont {Yuasa}},\ }\bibfield  {title} {\bibinfo {title} {Nonlinear
  {behavior} and {mode} {coupling} in {spin}-{transfer} {nano}-{oscillators}},\
  }\href {https://doi.org/10.1103/PhysRevApplied.2.061001} {\bibfield
  {journal} {\bibinfo  {journal} {Physical Review Applied}\ }\textbf {\bibinfo
  {volume} {2}},\ \bibinfo {pages} {061001(R)} (\bibinfo {year}
  {2014})}\BibitemShut {NoStop}%
\bibitem [{\citenamefont {Kim}(2012)}]{kim_spin-torque_2012}%
  \BibitemOpen
  \bibfield  {author} {\bibinfo {author} {\bibfnamefont {J.-V.}\ \bibnamefont
  {Kim}},\ }\bibfield  {title} {\bibinfo {title} {Spin-{torque}
  {oscillators}},\ }in\ \href
  {https://doi.org/10.1016/B978-0-12-397028-2.00004-7} {\emph {\bibinfo
  {booktitle} {Solid {State} {Physics}}}},\ Vol.~\bibinfo {volume} {63},\
  \bibinfo {editor} {edited by\ \bibinfo {editor} {\bibfnamefont {R.~L.}\
  \bibnamefont {Stamps}}\ and\ \bibinfo {editor} {\bibfnamefont {R.~E.}\
  \bibnamefont {Camley}}}\ (\bibinfo  {publisher} {Academic Press},\ \bibinfo
  {address} {San Diego},\ \bibinfo {year} {2012})\ pp.\ \bibinfo {pages}
  {217--294}\BibitemShut {NoStop}%
\bibitem [{\citenamefont {Petit-Watelot}\ \emph
  {et~al.}(2012{\natexlab{b}})\citenamefont {Petit-Watelot}, \citenamefont
  {Otxoa},\ and\ \citenamefont {Manfrini}}]{petit-watelot_electrical_2012}%
  \BibitemOpen
  \bibfield  {author} {\bibinfo {author} {\bibfnamefont {S.}~\bibnamefont
  {Petit-Watelot}}, \bibinfo {author} {\bibfnamefont {R.~M.}\ \bibnamefont
  {Otxoa}},\ and\ \bibinfo {author} {\bibfnamefont {M.}~\bibnamefont
  {Manfrini}},\ }\bibfield  {title} {\bibinfo {title} {Electrical properties of
  magnetic nanocontact devices computed using finite-element simulations},\
  }\href {https://doi.org/10.1063/1.3687915} {\bibfield  {journal} {\bibinfo
  {journal} {Applied Physics Letters}\ }\textbf {\bibinfo {volume} {100}},\
  \bibinfo {pages} {083507} (\bibinfo {year} {2012}{\natexlab{b}})}\BibitemShut
  {NoStop}%
\end{thebibliography}%

\end{document}